\documentclass[final,5p,times]{elsarticle}

\journal{International Journal of Electrical Power \& Energy Systems}

\usepackage{amsfonts,amsthm,amsmath,amssymb}
\usepackage{booktabs,pifont,makecell} 
\usepackage{enumitem}
\setlist[itemize]{itemsep=0.25em, parsep=0pt}
\setlist[enumerate]{itemsep=0.25em, parsep=0pt}
\usepackage{microtype}
\usepackage[colorlinks=true,
            citecolor=blue,     
            linkcolor=blue,     
            urlcolor=blue]      
            {hyperref}
\usepackage{tabularx} 
\usepackage{calc}
\usepackage{algorithm}  
\usepackage{algpseudocode}
\usepackage{multirow}
\usepackage{url}
\usepackage{graphicx}
\usepackage{subcaption}

\newcommand{\mt}[1]{\text{#1}} 
\newcommand{\refig}[1]{Fig.~\ref{#1}}
\newlength{\symbolwidth}
\newcommand{\nomitem}[2]{%
 #1 \> \parbox[t]{\linewidth-\symbolwidth}{#2.} \\%
}

\begin{document}

\begin{frontmatter}

\title{Stochastic Long-Term Joint Decarbonization Planning for Power Systems and Data Centers: A Case Study in PJM}

\author[ucr]{Zhentong Shao}
\author[ucr]{Nanpeng Yu\corref{cor1}}
\ead{nyu@ece.ucr.edu}
\author[ucr]{Daniel Wong}
\cortext[cor1]{Corresponding author}
\affiliation[ucr]{organization={Department of Electrical and Computer Engineering},
                  addressline={University of California, Riverside},
                  city={Riverside},
                  state={California},
                  postcode={92521},
                  country={USA}}

\begin{abstract}
    With the rapid growth of artificial intelligence (AI) and cloud services, data centers have become critical infrastructures driving digital economies, with increasing energy demand heightening concerns over electricity use and carbon emissions, emphasizing the need for carbon-aware infrastructure planning. Most studies assume static power systems, focus only on operational emissions, and overlook co-optimization. This paper proposes a dynamic joint planning framework that co-optimizes long-term data center and power system development over 15 years. The model determines siting, capacity, and type of data centers alongside power generation expansion, storage deployment, and retirements, accounting for both operational and embodied emissions. To handle multi-scale uncertainty, a large-scale two-stage stochastic program is formulated and solved via an enhanced Benders decomposition. Applied to the PJM Interconnection, with curated datasets released on GitHub, results show the system can support up to 55 GW peak data center demand, with Virginia (DOM) and Northern Illinois (ComEd) as optimal hosts. Compared to non-joint planning, the framework cuts investment cost by 12.6\%, operational cost by 8.25\%, and emissions by 5.63\%. Including lifecycle emissions further raises renewable deployment by 25.5\%, highlighting embodied carbon’s role in deeper decarbonization.
\end{abstract}


\end{frontmatter}

\section{Introduction}
\subsection{Motivation} 

Data centers have become critical infrastructures of the digital economy, enabling emerging technologies such as artificial intelligence (AI) \cite{INTRO-2}. Rapidly increasing energy demand from data centers has intensified concerns over carbon emissions \cite{INTRO-1}. While traditional efforts, such as improving power usage effectiveness (PUE) \cite{INTRO-4}, help reduce operational consumption, they fall short of meeting long-term decarbonization goals set by policymakers and environmental agencies \cite{INTRO-3}.

At the same time, data centers are tightly coupled with power systems, which themselves face renewable intermittency, shifting demand, and aging infrastructure \cite{INTRO-5}. However, as highlighted in the literature review (See Table~\ref{tab-1}), most existing studies adopt static power system assumptions, focus on single-site planning, or limit attention to operational emissions. Such approaches overlook embodied emissions \cite{dc-ijepes-1}, geographical efficiency variations \cite{dc-ijepes-2}, and the benefits of coordinated infrastructure development over time \cite{dc-ijepes-3}.

To overcome these limitations, this paper proposes a dynamic joint planning framework that co-optimizes data center siting, capacity, and workload dispatch with long-term power system expansion over a 15-year horizon. Unlike prior studies, our model is implemented on the real-world PJM system and incorporates both operational and embodied carbon emissions \cite{INTRO-7}, enabling a lifecycle perspective. The proposed framework provides actionable insights for a broad range of stakeholders engaged in power system planning and regulation. For Independent System Operators (ISOs), it generates quantitative signals to inform long-term capacity market auctions and to guide data center siting decisions toward grid-optimal locations. For policy makers, the results demonstrate how policy instruments—such as carbon pricing—can be strategically adjusted to balance economic efficiency and environmental sustainability in data center development. Furthermore, the framework assists public utility commissions in assessing and approving proactive grid expansion plans that ensure sufficient capacity to meet the rapidly increasing electricity demand from large-scale data centers.

Furthermore, the limitations of separate planning extend beyond methodological gaps to tangible system risks. Uncoordinated data center growth often leads to unforeseen transmission congestion and localized overloads, forcing utilities to undertake costly, reactive upgrades that increase total social cost. In contrast, a proactive co-optimization strategy—as proposed in this work—coordinates siting and capacity decisions with grid expansion, thereby minimizing congestion risks, avoiding inefficient retrofitting, and enhancing long-term reliability while aligning decarbonization goals with infrastructure development.

\subsection{Literature review} 

\begin{table*}[!t]
\centering
\caption{Comparison of representative literature on data center planning}
\vspace{-0.5em}
\label{tab-1}
\begin{tabular}{@{}lcccclcl@{}}
\toprule
\textbf{Reference} & \textbf{Joint} & \textbf{Retirement} & \textbf{Practical} & \textbf{Carbon} & \textbf{Planning Scope} & \textbf{Planning } & \textbf{Planning Horizon} \\
                   & \textbf{Planning} & \textbf{Planning} & \textbf{Case Study} &\textbf{Emissions} & \textbf{ } & \textbf{Structure} & \textbf{ } \\
\midrule
\cite{R1-google}     & \ding{55}  & \ding{55} & \ding{55} & OP         & Servers             & One-shot  & Short-term (1 yr)   \\
\cite{R2-nationwide} & \ding{55}  & \ding{55} & \ding{55} & OP         & Data Center         & One-shot  & Mid-term (5 yrs)    \\
\cite{R3-XJTU}       & \ding{55}  & \ding{55} & \ding{55} & OP         & Microgrid           & One-shot  & Short-term (1 yr)   \\
\cite{R4}            & \ding{55}  & \ding{55} & \ding{55} & \ding{55}  & Data Center         & One-shot  & Short-term (1 yr)   \\
\cite{R5}            & \ding{55}  & \ding{55} & \ding{55} & \ding{55}  & Microgrid           & One-shot  & Mid-term (4 yrs)    \\
\cite{R6}            & \ding{55}  & \ding{55} & \ding{55} & \ding{55}  & Data Center         & Dynamic   & Long-term (20 yrs)  \\
\cite{R7}            & \ding{55}  & \ding{55} & \ding{55} & OP         & Data Center         & One-shot  & Short-term (1 yr)   \\
\cite{R8}            & \ding{55}  & \ding{55} & \ding{55} & \ding{55}  & Microgrid           & One-shot  & Short-term (1 yr)   \\
\cite{R9}            & \ding{55}  & \ding{55} & \ding{55} & OP         & Building System     & One-shot  & Short-term (1 yr)   \\
\textbf{This paper}  & \checkmark & \checkmark& \checkmark& OP \& EM   & \makecell{Power System \& DC}  & Dynamic   & Long-term (15 yrs)  \\
\bottomrule
\end{tabular}
\\[0.5em]
\raggedright \textit{Notes:} \\
\textit{$\diamond$ Joint planning makes investment decisions for both the power system and data centers, enabling coordinated development. In contrast, existing studies assume a fixed power system and plan data centers accordingly.} \\
\textit{$\diamond\diamond$ OP refers to operational carbon emissions, OP \& EM refers to both operational and embodied carbon emissions.} \\
\textit{$\diamond\diamond\diamond$ Dynamic planning involves annual decisions over time, while One-shot determines all decisions at once.} 
\end{table*}

The existing literature on data center planning can be broadly classified into two categories. The first category focuses on individual data centers, analyzing expansion planning and energy management from a single-site perspective \cite{OPR-1, OPR-2, OPR-3, OPR-4, OPR-5, R1-google}. These studies typically neglect interactions with the surrounding power system. The second category considers the coordination between data centers and power systems, but often in a limited manner, either through simplified assumptions or partial integration of power system factors. Our paper belongs to this second category, adopting the perspective of regional agencies such as public utilities commissions or ISOs. We investigate how joint planning of data centers and power systems can enhance long-term societal benefits through improved energy use and carbon reduction.

Compared to single-site planning, joint planning is more complex. Multi-site data center planning must capture system-wide power balance and flows \cite{OPR-6}, while accounting for renewable supply uncertainty and transmission congestion \cite{OPR-7}. Hence, the joint planning problem becomes a non-linear combinatorial optimization, posing computational challenges for practical grids \cite{OPR-8}. For single-site data centers, such complexities are often ignored, yielding much simpler models \cite{OPR-3}.

Geographical impacts further complicate joint planning by affecting efficiency and emissions \cite{R2-nationwide}. Regional climate differences influence cooling demand \cite{OPR-9}, which constitutes a major share of data center energy use. Embodied carbon emissions from construction also vary by location \cite{OPR-10}. Assessing these emissions is critical for sustainable planning. In addition, resource endowment shapes renewable feasibility and achievable decarbonization. Although \cite{R2-nationwide, R4, R6, R7} include power system factors, key gaps remain. For example, \cite{R4, R6, R7} omit regional energy efficiency and carbon factors, as well as expansion planning. While \cite{R2-nationwide} includes comprehensive geography, it simplifies the power system into preset locational marginal prices (LMP) and carbon emissions (LMCE). Moreover, most studies consider only operational emissions, overlooking embodied ones essential for long-term planning.

The modeling granularity of data centers also affects computational efficiency and accuracy. Capturing server racks, cooling technologies, and workload transfer ensures realistic energy representation. Yet detailed models (e.g., CPU utilization and QoS control \cite{R1-google, R4, R7, R8}) are too complex for long-term horizons, while oversimplified models cause inaccuracies. For instance, \cite{R3-XJTU} treats data centers as load curves without cooling, while \cite{R6, R9} reduce them to shiftable loads. By contrast, rack-level modular modeling \cite{R2-nationwide,R5} effectively captures major components, balancing accuracy and tractability for long-term carbon-aware planning.

Multi-source uncertainties in joint planning remain underexplored. Many studies \cite{R3-XJTU,R4,R5,R6,R7,R8,R9} focus on short-term issues such as renewable intermittency and demand variation. For instance, \cite{R6} applies scenario-based stochastic optimization (SO) to handle renewable and load uncertainty, and \cite{R7} considers demand-side flexibility under electricity–carbon markets. \cite{R8} proposes a three-stage distributionally robust optimization (DRO) for regional electricity–heat systems with wind uncertainty. Yet long-term uncertainties in demand and workload growth, crucial for joint planning, are often neglected \cite{R2-nationwide,R9}. Practical limits in handling large scenario sets also reduce accuracy and reliability. These gaps highlight the need for advanced modeling and computational acceleration techniques for effective joint planning.

Table~\ref{tab-1} compares representative studies, emphasizing gaps addressed by this work. First, most of the existing work assume fixed power systems, optimizing only data center siting and missing the benefits of joint planning. As shown, none co-optimize investments in both systems, overlooking coordinated development under rising demand. Second, while some papers consider operational (OP) emissions, embodied emissions (EM) are rarely included, limiting lifecycle assessments. Third, most adopt static planning, ignoring evolving infrastructure. Although \cite{R6} applies dynamic planning, it lacks embodied carbon and joint expansion. In contrast, our work proposes a dynamic joint planning framework over 15 years for the PJM system, co-optimizing power system and data center investments while incorporating both operational and embodied emissions. This comprehensive approach bridges key gaps and offers a realistic, carbon-aware pathway for long-term infrastructure planning.

\subsection{Contributions} 

This paper addresses key methodological gaps in the existing literature by proposing a dynamic joint planning framework that integrates the long-term development of power systems and data centers. The main contributions are summarized:

\begin{enumerate}[leftmargin=*]
    \item \textbf{Dynamic joint planning of coupled infrastructures:} 
    Unlike prior studies that optimize data center deployment with static power system assumptions, this work co-optimizes long-term investments in both data centers and power generation assets over a 15-year horizon. The proposed model captures spatial and temporal interdependencies between digital demand and power supply, enabling coordinated infrastructure development.
    \item \textbf{Lifecycle carbon integration:} 
    This work is among the first to integrate both operational and embodied carbon emissions into joint infrastructure planning. By accounting for emissions from facility construction, equipment manufacturing, and long-term operations, the framework supports a more comprehensive and carbon-conscious planning pathway. Incorporating embodied carbon also incentivizes renewable energy development, thereby enabling deeper decarbonization.
    \item \textbf{Real-world application under multi-scale uncertainty:} 
    The framework is applied to the PJM interconnection, one of the largest power systems in North America. The case study simultaneously addresses long-term demand growth and short-term renewable intermittency. To improve tractability, the model is solved using Benders decomposition enhanced for faster convergence. All processed datasets and system parameters are curated and publicly released via GitHub \cite{github-pjm-case}, providing a valuable resource for future research on integrated infrastructure planning.
\end{enumerate}

\vspace{-0.5em}
In addition to methodological contributions, the case study conducted on the PJM system yields several empirical insights that demonstrate the practical value of the proposed framework:  
\begin{itemize}[leftmargin=*]
    \item The results show that the PJM system can accommodate up to 55 GW of peak data center demand, with the DOM (Virginia) and ComEd (Northern Illinois) zones identified as the most suitable hosting regions.  
    \item Joint planning of power systems and data centers leads to notable system-wide benefits. Compared to independent data center planning, it reduces total investment costs by 12.63\%, operational costs by 8.25\%, and carbon emissions by 5.63\%.  
    \item Compared to the case where embodied carbon emissions are not considered, incorporating embodied carbon into the planning framework leads to a 25.5\% increase in renewable capacity deployment and a 16.9\% reduction in operational carbon emissions.
\end{itemize}

\vspace{-1em}
\subsection{Paper organization}
The remainder of this paper is organized as below. Section II provides the planning framework and the mathematical formulation of the joint planning model. Section III derives the solution strategy and illustrates the solution approach of the strengthened Benders decomposition. Section IV shows the results of numerical studies on the PJM interconnection. Finally, Section V concludes the paper.

\vspace{-1em}
\section{Two-stage joint planning framework}
\vspace{-0.5em}
The joint data center and power system planning model is constructed from the perspective of a regional government agency, such as public utility commission, whose objective is to satisfy computing and electric load in a way that reduces both costs and greenhouse gas emissions. The optimal planning outcome can guide both data center and energy resource developers to strategically locate data centers and power plants in a manner that maximizes system-wide benefits.

\begin{figure}[tbp]
    \centering
    \includegraphics[width=3.5in]{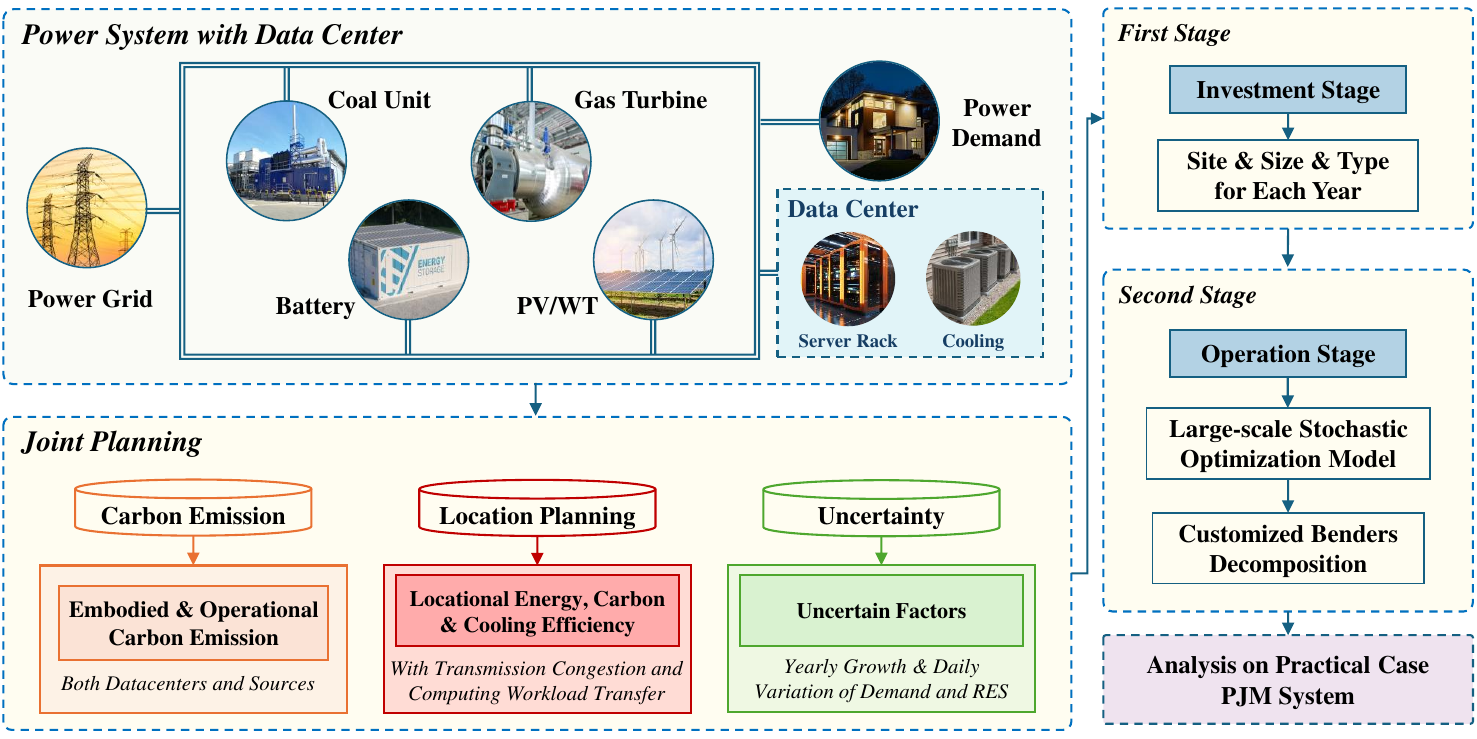}
    \vspace{-1.5em}
    \caption{Structure of joint planning of power systems with data centers.}
    \label{fig1}
\end{figure}

\begin{figure}[tbp]
    \centering
    \includegraphics[width=3.5in]{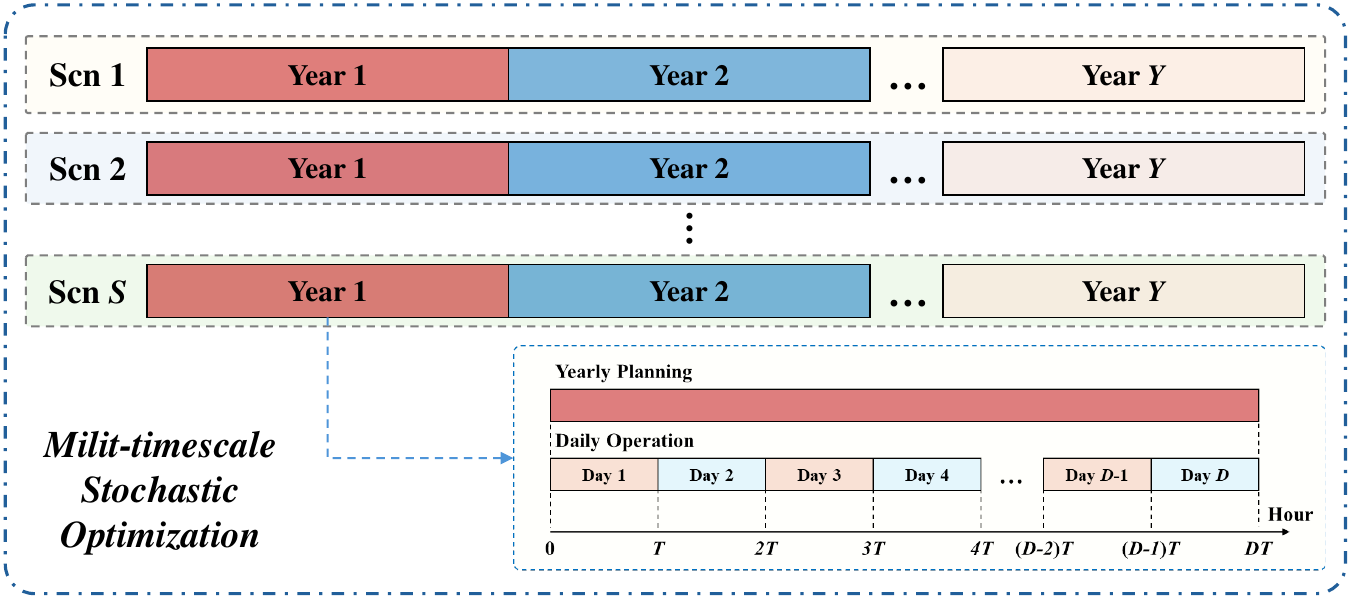}
    \vspace{-1.5em}
    \caption{Illustration of multi-timescale stochastic planning optimization.}
    \label{fig2}
    \vspace{-1.5em}
\end{figure}

This paper proposes a two-stage joint planning framework to determine the optimal siting, sizing, and cooling system selection of data center and power system resources to achieve decarbonization and investments minimization. The proposed framework models a two-stage decision-making process. In the first stage, long-term investment decisions are made to determine the placement, capacity, and type of resources. The second stage focuses on verifying the operational feasibility of the system by analyzing short-term daily operations. The overall structure of the joint planning framework is shown in \refig{fig1}. The multi-timescale stochastic planning timeline is illustrated in \refig{fig2}, indicating that the proposed model incorporates a large-scale framework by considering multiple years and expanding to a large number of scenarios. This presents a significant computational challenge, which will be addressed in the next section by introducing a Benders decomposition method. 

Key planning settings for power systems and data centers are given as below:

\vspace{-0.5em}
\begin{enumerate}[leftmargin=*]
    \item[a)] \textit{Data center expansion:} The planning stage determines location, size (number of server racks), and cooling options (represented by locational power utilization efficiency). Decisions are annual, accounting for embodied carbon and minimizing investment costs. In operation, the goal is to reduce O\&M costs and operational emissions by distributing data center power loads across sites, dispatching active servers, and enabling demand response. This considers locational differences in energy prices, and carbon intensity.
    
    \item[b)] \textit{Power system expansion:} Planning focuses on annual renewable energy resources (RES) and natural gas-fueled power plants (NG) development and coal-powered power plants (CG) retirements to meet carbon targets. Decisions include site selection, capacity, and construction scheduling. In operation, the system is modeled with multi-period optimal power flow under transmission limits, capturing locational differences and congestion. The model incorporates annual load growth from both system demand and data centers, and evaluates reliability under short-term load/renewable uncertainties and long-term data center demand growth.
\end{enumerate}

\begin{table}[tbp]
\centering
\caption{Summary of model indices}
\vspace{-0.5em}
\label{tab-2}
\begin{tabularx}{\linewidth}{lX}
\toprule
\textbf{Index} & \textbf{Description} \\
\midrule
$ d $   & Index of days, $d \in \mathcal{N}_{d}$ \\ 
$ i,j $ & Index of power system nodes/locations, $i,j \in \mathcal{N}_{i}$ \\ 
$ ij $  & Index of transmission lines from node $i$ to node $j$, $ij \in \mathcal{N}_{l}$ \\ 
$ k $   & Index of candidate generation resources, $k \in \mathcal{N}_{k}$ \\ 
$ l $   & Index of transmission lines, $l \in \mathcal{N}_{l}$ \\ 
$ m $   & Index of server types, $m \in \mathcal{N}_{m}$ \\ 
$ t $   & Index of hours, $t \in \mathcal{N}_{t}$ \\ 
$ y $   & Index of years, $y \in \mathcal{N}_{y}$ \\ 
\bottomrule
\end{tabularx}
\end{table}

\vspace{-1.5em}
\section{Long-term joint decarbonization planning model}
The joint planning model seeks to simultaneously minimize both the total system costs (TC) and carbon emissions (TE) across the investment and operational stages. For clarity, the model indices are summarized in Table~\ref{tab-2}, while other nomenclatures are provided in the Appendix.

\vspace{-1em}
\subsection{Economic objective} 
The economic objective is defined in \eqref{obj-eco}, encompassing the investment cost in the planning stage, represented by $ F(\mathbf{z}) $, as well as the operational cost in the operational stage represented by $ G(\mathbf{x}) $, where $\mathbf{z}$ and $\mathbf{x}$ denote the sets of decision variables for the planning and operational stages, respectively. 
\begin{align}
    \mt{TC} = \min_{\mathbf{z} \in \mathbb{Z},~\mathbf{x} \in \mathbb{X}} \left[
    \underset{\mt{Planning Stage}}{F(\mathbf{z})} + \underset{\mt{Operational Stage}}{G(\mathbf{x})}
    \right] \label{obj-eco}
\end{align}

The planning investment $F(\mathbf{z})$ calculated in \eqref{obj-e-1} uses the investment discount factor $\sigma$. 
Here, $C_{\mt{GR}}$, $C_{\mt{DC}}$, $C_{\mt{TL}}$, and $C_{\mt{TC}}$ are the investment costs in generation resources, data centers, transmission lines, and data transmission facilities, respectively; 
$C_{\mt{RT}}$ is the disposal value of retired resources, and $\mt{RV}$ is the total residual value at the end of the planning horizon.
\begin{equation}
    F(\mathbf{z}) = \sum_{y \in \mathcal{N}_{y}} \left[ 
         \frac{C_{\mt{GR}}^{y} + C_{\mt{DC}}^{y}  + C_{\mt{TL}}^{y} + C_{\mt{TC}}^{y} - C_{\mt{RT}}^{y}}{(1+\sigma)^{y}}
    \right] - \frac{\mt{RV}}{(1+\sigma)^{|\mathcal{N}_{y}|}}\label{obj-e-1}
\end{equation}

The operational cost $G(\mathbf{x})$ is defined in \eqref{obj-o-1}, where 
$C_{\mt{gen}}^{y}$, $C_{\mt{om}}^{y}$, $C_{\mt{ex}}^{y}$, and $C_{\mt{curt}}^{y}$ 
represent the generation cost, maintenance cost, energy transaction cost, and 
RES curtailment cost, respectively.
\begin{equation}
    G(\mathbf{x}) = \sum_{y \in \mathcal{N}_{y}} \frac{
    C_{\mt{gen}}^{y} + C_{\mt{om}}^{y} + C_{\mt{ex}}^{y}  + C_{\mt{curt}}^{y}
    }{(1+\sigma)^{y}} \label{obj-o-1}
\end{equation}

\vspace{-1em}
\subsection{Planning investments $ F(\mathbf{z}) $} 

The terms $C_{\mt{GR}}$, $C_{\mt{DC}}$, $C_{\mt{TL}}$, and $C_{\mt{TC}}$ in \eqref{obj-e-1} are calculated in formulation \eqref{cal-Fz}. Specifically, equation \eqref{obj-e-2} denotes planning costs for generation resources, including natural gas (NG), solar (PV), wind turbines (WT) and energy storage (ES) units. $\xi^{\mt{inv}}$ denotes the unit capital cost, while $\mt{IC}_{k}^{y,i}$ is the installed capacity of resource $k$ at year $y$ and node $i$. Equation \eqref{obj-e-3} captures data center planning costs, including server investments and fixed resources (buildings, land, materials), proportional to installed server racks ($\mt{IN}_{\mt{DC}}$). Index $m$ in \eqref{obj-e-3} encodes server type selection. Equation \eqref{obj-e-4} quantifies transmission line expansion costs. Equation \eqref{obj-e-5} models data transmission facility expansion costs. 

\vspace{-0.5em}
\begin{subequations}
    \begin{align}
    C_{\mt{GR}}^{y} &= \sum_{i \in \mathcal{N}_{i}}
        \sum_{k \in \{\mt{NG,PV,WT,ES}\}} \xi^{\mt{inv}}_{k} \left( \mt{IC}_{k}^{y,i} - \mt{IC}_{k}^{y-1,i} \right)
    \label{obj-e-2}
    \\
    C_{\mt{DC}}^{y} & = \sum_{i \in \mathcal{N}_{i}} 
    \left[ 
    \begin{aligned}
        &\sum_{m \in \mathcal{N}_{m}} \nolimits \left( \gamma_{\mt{tech}}^{y} \xi^{\mt{inv}}_{\mt{SRV,m}} \mt{BN}_{\mt{SRV,m}}^{y,i} \right)
        \\
        &\hspace{7em}+
        \xi^{\mt{inv}}_{\mt{Fix} } \left( \mt{IN}_{\mt{DC}}^{y,i} -  \mt{IN}_{\mt{DC}}^{y-1,i} \right)
    \end{aligned} 
    \right] 
    \label{obj-e-3}
    \\
    C_{\mt{TL}}^{y} &= \sum_{l \in \mathcal{N}_{l}} H_{l}~ \xi^{\mt{inv}}_{\mt{TL}}  \left( \mt{IC}_{\mt{TL}}^{y,l} - \mt{IC}_{\mt{TL}}^{y-1,l} \right) \label{obj-e-4}
    \\
    C_{\mt{TC}}^{y} &= \sum_{i \in \mathcal{N}_{i}} \sum_{j \in \mathcal{N}_{i}, j \neq i } Z_{ij} H_{ij} \xi^{\mt{inv}}_{\mt{TC}} \left( \bar{Y}_{ij}^{y} - \bar{Y}_{ij}^{y-1} \right) \label{obj-e-5}
\end{align}
\label{cal-Fz}
\end{subequations}
\vspace{-0.5em}

In formulation \eqref{obj-e-5}, the parameter $Z_{ij}$ indicates whether node $i$ can serve node $j$ based on data propagation delay, and it is calculated via \eqref{cal-Z-value} before optimization. Specifically, if $H_{ij} \leq H_{0}$, node $i$ can serve node $j$ within the delay limit; otherwise, it cannot.
Equation \eqref{st-p-dc-4} sets the maximum acceptable distance $H_{0}$, where $T_{\mt{req}}$ is the delay threshold and $\tau_{0}$ the per-unit distance delay. Network parameters are fixed, with a typical delay of 0.82 ms per 100 miles \cite{delay}.

\vspace{-1em}
\begin{subequations}
    \begin{align}
        &Z_{ij, i \neq j} = \begin{cases}
                1,~H_{ij} \leq H_{0}
                \\
                0,~H_{ij} > H_{0}
            \end{cases} \label{st-p-dc-3}
        \\
        &H_{0} = T_{\mt{req}} / \tau_{0} \label{st-p-dc-4}
    \end{align}
    \label{cal-Z-value}
\end{subequations}
\vspace{-1em}

Equation \eqref{obj-e-6} calculates the disposal value of coal-fired generators (CG) and data center servers (SRV). CG units are restricted to retirement, with no new installations allowed to align with decarbonization objectives. Note that the factor $R_{\mt{CG}}^{y}$ in \eqref{obj-e-6} evaluates the residual value of the retired CG units. In this paper, we set $R^{y}_{\mt{CG}} = 80\%$ in the first year, and then linearly decrease it by 4.5\% each year thereafter. 
\begin{equation}
    \mt{C}_{\mt{RT}}^{y} = \sum_{i \in \mathcal{N}_{i}} \left[
    \begin{aligned}
        &R^{y}_{\mt{CG}} \xi^{\mt{inv}}_{\mt{CG}} \left( \mt{IC}_{\mt{CG}}^{y-1,i} - \mt{IC}_{\mt{CG}}^{y,i} \right) 
        \\ 
        &\hspace{2em}+ 
        \sum_{m \in \mathcal{N}_{m}} \nolimits \left( R^{0}_{\mt{SRV},m} \gamma_{\mt{tech}}^{y} \xi^{\mt{inv}}_{\mt{SRV,m}} \mt{RN}_{\mt{SRV,m}}^{y,i} \right)
    \end{aligned}
    \right] \label{obj-e-6}
\end{equation}

Equation \eqref{obj-e-7} computes the residual value of all resources at the end of the planing horizon, and the residual value rate of resources installed in $y$-th year ($R_{k}^{y}$) is computed in equation \eqref{obj-e-8}.

\vspace{-1em}
\begin{align}
    \mt{RV} &= \sum_{y \in \mathcal{N}_{y}} \left[
        \begin{aligned}
            &\sum_{i \in \mathcal{N}_{i}} \sum_{k \in \{\mt{NG,PV,WT,ES}\}} R^{y}_{k} \xi^{\mt{inv}}_{k} \left( \mt{IC}_{k}^{y,i} - \mt{IC}_{k}^{y-1,i} \right) 
            \\
            &\hspace{0.5em}+ 
            \sum_{i \in \mathcal{N}_{i}} R^{y}_{\mt{Fix}} \xi^{\mt{inv}}_{\mt{Fix} } \left( \mt{IN}_{\mt{DC}}^{y,i} -  \mt{IN}_{\mt{DC}}^{y-1,i} \right)
            \\
            &\hspace{1.5em}+\sum_{l \in \mathcal{N}_{l}} R^{y}_{\mt{TL}} H_{l} \xi^{\mt{inv}}_{\mt{TL}} \left( \mt{IC}_{\mt{TL}}^{y,l} - \mt{IC}_{\mt{TL}}^{y-1,l} \right) 
            \\
            &\hspace{2.5em}+ 
            \sum_{i \in \mathcal{N}_{i}} \sum_{j \in \mathcal{N}_{i}, j \neq i } R^{y}_{\mt{TC}} \xi^{\mt{inv}}_{\mt{TC}} Z_{ij} H_{ij} \left( \bar{Y}_{ij}^{y} - \bar{Y}_{ij}^{y-1} \right)
        \end{aligned}
    \right] \label{obj-e-7}
    \\
    R^{y}_{k} &= \frac{L_{k}-(|\mathcal{N}_{y}|-y+1)}{L_{k}}(1-R^{0}_{k}) + R^{0}_{k}, \nonumber
    \\
    & \hspace{7em} \forall k \in \{\mt{NG,PV,WT,ES,Fix,TC,TL}\} 
    \label{obj-e-8}
\end{align}
\vspace{-1em}

\vspace{-2em}
\subsection{Operational costs $ G(\mathbf{x}) $} 

The terms $C_{\mt{gen}}^{y}$, $C^{y}_{\mt{om}}$, $C_{\mt{ex}}^{y}$ and $C_{\mt{curt}}^{y}$ in \eqref{obj-o-1} are calculated in
formulation \eqref{cal-Gx}. Specifically, electricity generation costs ($C_{\mt{gen}}$) are calculated in equation \eqref{obj-o-2} for NG and CG units. The maintenance costs ($C_{\mt{om}}$) for all generation resources are given in \eqref{obj-o-3}.  The power transactions ($C_{\mt{ex}}$) between the PJM and other ISOs are specified in \eqref{obj-o-4}. The RES curtailments ($C_{\mt{curt}}$) are punished in \eqref{obj-o-5}.

\vspace{-1em}
\begin{subequations}
    \begin{align}
    C_{\mt{gen}}^{y} &= \sum_{d \in \mathcal{N}_{d}} \sum_{t \in \mathcal{N}_{t}} \sum_{i \in \mathcal{N}_{i}} \left(\xi_{\mt{NG}}^{gen}P_{\mt{NG}}^{y,d,t,i} \Delta t + \xi_{\mt{CG}}^{gen} P_{\mt{CG}}^{y,d,t,i} \Delta t\right) \label{obj-o-2}
    \\
    C^{y}_{\mt{om}} &= \sum_{d \in \mathcal{N}_{d}} \sum_{t \in \mathcal{N}_{t}} \sum_{i \in \mathcal{N}_{i}} 
    \left[ 
    \begin{aligned}
        &\sum_{k \in \{\mt{NG,CG,PV,WT}\}} \xi_{k}^{om} P_{k}^{y,d,t,i} \Delta t
        \\
        &\hspace{3em}+\xi_{\mt{ES}}^{om} \left( P_{\mt{ESD}}^{y,d,t,i} \Delta t + P_{\mt{ESC}}^{y,d,t,i} \Delta t \right)
    \end{aligned}
    \right]
    \label{obj-o-3}
    \\
    C_{\mt{ex}}^{y} &= \sum_{d \in \mathcal{N}_{d}} \sum_{t \in \mathcal{N}_{t}} \sum_{i \in \mathcal{N}_{i}} \left(\xi^{buy}_{i,t} P_{\mt{BUY}}^{y,d,t,i} \Delta t - \xi^{sell}_{i,t} P_{\mt{SELL}}^{y,d,t,i} \Delta t \right) 
    \label{obj-o-4}
    \\
    C_{\mt{curt}}^{y} &= \sum_{d \in \mathcal{N}_{d}} \sum_{t \in \mathcal{N}_{t}} \sum_{i \in \mathcal{N}_{i}} \left[
    \begin{aligned}
        &\xi^{curt}_{\mt{WT}} \left(P_{\mt{WT,cap}}^{y,d,t,i} - P_{\mt{WT}}^{y,d,t,i}\right) \Delta t
        \\
        &\hspace{3em}+\xi^{curt}_{\mt{PV}} \left(P_{\mt{PV,cap}}^{y,d,t,i} - P_{\mt{PV}}^{y,d,t,i}\right) \Delta t
    \end{aligned}
     \right]  \label{obj-o-5}
\end{align}
\label{cal-Gx}
\end{subequations}
\vspace{-1em}

\vspace{-1em}
\subsection{Environmental objective} 
To minimize the system-wide carbon emissions, including embodied carbon emissions $\mt{CE}_{\mt{EMB}}$ from the planning stage and operational carbon emissions $\mt{CE}_{\mt{OPR}}$ from the operational stage. Specifically, equation \eqref{obj-em-1} summarizes the carbon emissions. Equation \eqref{obj-em-2} calculates the total embodied carbon emissions during system construction, and equation \eqref{obj-em-3} represents the operational carbon emissions from electricity generation.

\vspace{-1em}
\begin{subequations}
\begin{align}
    \mt{TE} &= \min_{\mathbf{z} \in \mathbb{Z},~\mathbf{x} \in \mathbb{X}} \sum_{y \in \mathcal{N}_{y}} \left[
    \underset{\mt{Planning Stage}}{\mt{CE}_{\mt{EMB}}^{y}} + \underset{\mt{Operational Stage}}{\mt{CE}_{\mt{OPR}}^{y}}
    \right] \label{obj-em-1}
    \\
    \mt{CE}_{\mt{EMB}}^{y} &= \sum_{i \in \mathcal{N}_{i}} \left[ 
        \begin{aligned}
            &\sum_{k \in \{\mt{NG,PV,WT,ES}\}} \chi^{\mt{emb}}_{k} \left( \mt{IC}_{k}^{y,i} - \mt{IC}_{k}^{y-1,i} \right)
            \\
            &
            +\sum_{j \in \mathcal{N}_{i}, j \neq i } \chi^{\mt{emb}}_{\mt{TC}} Z_{ij} H_{ij} \left( \bar{Y}_{ij}^{y} - \bar{Y}_{ij}^{y-1} \right)
            \\
            &
            +\sum_{ m \in \mathcal{N}_{m}} \lambda^{y}_{\mt{tech}} \chi^{\mt{emb}}_{\mt{SRV,m}} \left( \mt{BN}_{\mt{SRV},m}^{y,i} \right) 
        \end{aligned}
    \right] +
    \nonumber
\end{align}
\begin{align}
    & \hspace{-0.75em} \sum_{i \in \mathcal{N}_{i}} \chi^{\mt{emb}}_{\mt{Fix}} \left( \mt{IN}_{\mt{DC}}^{y,i} 
    - \mt{IN}_{\mt{DC}}^{y-1,i} \right) + \sum_{l \in \mathcal{N}_{l}} \chi^{\mt{emb}}_{\mt{TL}} H_{l} \left( \mt{IC}_{\mt{TL}}^{y,l} - \mt{IC}_{\mt{TL}}^{y-1,l} \right)     \label{obj-em-2}
    \\
    & \hspace{-0.5em} \mt{CE}_{\mt{OPR}}^{y} = \sum_{d \in \mathcal{N}_{d}} \sum_{t \in \mathcal{N}_{t}} \sum_{i \in \mathcal{N}_{i}} \left(\chi_{\mt{NG}}^{gen}P_{\mt{NG}}^{y,d,t,i} + \chi_{\mt{CG}}^{gen} P_{\mt{CG}}^{y,d,t,i} \right) \Delta t \label{obj-em-3}
\end{align}
\end{subequations}

\vspace{-1em}
\subsection{PJM system planning constraints} 
The planning constraints for the PJM system are given in \eqref{pjm-planning-constraints}, with each constraint applying $\forall y \in \mathcal{N}_{y}, \forall i \in \mathcal{N}_{i}$. Specifically, equations \eqref{st-p-1}-\eqref{st-p-4} indicate that the newly installed capacities of generation resources and transmission lines must follow discrete unit sizes. Notably, \eqref{st-p-4} specifies that coal-fired generators can only be retired. Thus, we have $\mt{IC}_{\mt{CG}}^{y-1,i} \geq \mt{IC}_{\mt{CG}}^{y,i}$. Constraints \eqref{st-p-5} limits and ensures the maximum retired capacity of CG units according to the scheduled retirement plan. Constraint \eqref{st-p-6} limits the installed capacity of power generation resources. Meanwhile, the renewable portfolio standards (RPS), which require regions to install RES to a certain degree by specific years \cite{RPS-2024PJM}, are also enforced in this constraint. Constraints \eqref{st-p-7} ensure the non-negativity and discreteness of the planning variables.

\vspace{-1em}
\begin{subequations}
\begin{align}
    &\mt{IC}_{k}^{y,i} = \mt{N}_{k}^{y,i} P_{k}^{\mt{unit}} + \mt{IC}_{k}^{y-1,i},~\forall k \in \{\mt{NG,PV,WT}\} \label{st-p-1}
    \\
    &\mt{IC}_{\mt{ES}}^{y,i} = \mt{N}_{\mt{ES}}^{y,i} E_{\mt{ES}}^{\mt{unit}} + \mt{IC}_{\mt{ES}}^{y-1,i} \label{st-p-2}
    \\
    &\mt{IC}_{\mt{TL}}^{y,l} = \mt{N}_{\mt{TL}}^{y,l} P_{\mt{TL}}^{\mt{unit}} + \mt{IC}_{\mt{TL}}^{y-1,l} \label{st-p-3}
    \\
    &\mt{IC}_{\mt{CG}}^{y-1,i} = \mt{N}_{\mt{CG}}^{y,i} P_{\mt{CG}}^{\mt{unit}} + \mt{IC}_{\mt{CG}}^{y,i} \label{st-p-4}
    \\
    &0 \leq \mt{IC}_{\mt{CG}}^{y,i},~~\sum_{i \in \mathcal{N}_{i}}  \nolimits \mt{N}_{\mt{CG}}^{y,i} P_{\mt{CG}}^{\mt{unit}} \leq \bar{\mt{RT}}_{\mt{CG}}^{y} \label{st-p-5}
    \\
    &\underline{\mt{IC}}_{k}^{y,i} \leq \mt{IC}_{k}^{y,i} \leq \bar{\mt{IC}}_{k}^{y,i},~\forall k \in \{\mt{NG,PV,WT,ES}\} \label{st-p-6}
    \\
    &\mt{N}_{k}^{y,i} \in \mathbb{Z}_{+},\forall k \in \{\mt{NG,PV,WT,ES,TL}\}; \mt{N}_{\mt{CG}}^{y,i} \in \mathbb{Z}_{+} \label{st-p-7}
\end{align}
\label{pjm-planning-constraints}
\end{subequations}
\vspace{-2em}

\subsection{Data center planning constraints} 
The constraints for data centers' planning are given in formulation \eqref{dc-plan} ($\forall y \in \mathcal{N}_{y}, \forall i \in \mathcal{N}_{i}, m \in \mathcal{N}_{m}$). Constraint \eqref{st-p-dc-1a} computes the number of installed servers from newly added ($\mt{BN}$) and retired ($\mt{RN}$) servers. Constraint \eqref{st-p-dc-1b1} ensures the number of servers in data centers does not decrease over time, while \eqref{st-p-dc-1b2} gives the total servers at node $i$.
Constraint \eqref{st-p-dc-2a} forces servers to retire once they reach their lifetime, and \eqref{st-p-dc-2b} restricts installed and retired server variables to discrete values.

\vspace{-1em}
\begin{subequations}
\begin{align}
    &\mt{IN}_{\mt{SRV,m}}^{y,i} - \mt{IN}_{\mt{SRV,m}}^{y-1,i} =  \mt{BN}_{\mt{SRV},m}^{y,i} - \mt{RN}_{\mt{SRV},m}^{y,i} \label{st-p-dc-1a}    
    \\
    &\mt{IN}_{\mt{SRV,m}}^{y,i} - \mt{IN}_{\mt{SRV,m}}^{y-1,i} \geq 0, \label{st-p-dc-1b1}
    \\
    &\mt{IN}_{\mt{DC}}^{y,i} = \sum_{m \in \mathcal{N}_{m}} \nolimits \mt{IN}_{\mt{SRV.m}}^{y,i} \label{st-p-dc-1b2}
    \\
    &\mt{RN}_{\mt{SRV},m}^{(y + L_{\mt{SVR,m}}), i} =\begin{cases}
        \mt{BN}_{\mt{SRV},m}^{y,i}, 
        & \text{if } y+L_{\mt{SVR,m}} \leq |N_{y}| \\
        0, 
        & \text{otherwise}
    \end{cases}  \label{st-p-dc-2a}
    \\
    &\mt{RN}_{\mt{SRV},m}^{y, i}, \mt{BN}_{\mt{SRV},m}^{y, i} \in \mathbb{Z}_{+} \label{st-p-dc-2b}
\end{align}
\label{dc-plan}
\end{subequations}
\vspace{-1em}

\vspace{-1em}
\subsection{Operational constraints for the PJM system} 
The operational constraints for the PJM power system are defined in formulations \eqref{pjm-opr-set-1}--\eqref{pjm-opr-set-3}, where each constraint applies for all $y \in \mathcal{N}_{y}$, $d \in \mathcal{N}_{d}$, $t \in \mathcal{N}_{t}$, and $i \in \mathcal{N}_{i}$.

The system-wide operational constraints are given in \eqref{pjm-opr-set-1}. Specifically, equation \eqref{st-p-a1} defines the nodal power generation. Equation \eqref{st-p-a2} enforces the system-wide power balance. Equation \eqref{st-p-a3} calculates the transmission line power flows using power transfer distribution factors ($\Gamma_{l,i}$). Constraints \eqref{st-p-a4} enforce power flow limits on transmission lines. Constraint \eqref{st-p-a5} imposes capacity limitations on power transactions through the system interface.

\vspace{-1em}
\begin{subequations}
\begin{align}
    &P^{y,d,t,i}_{\mt{GEN}} = \left[
        \begin{aligned}
            &P^{y,d,t,i}_{\mt{NG}} + P^{y,d,t,i}_{\mt{CG}} + P^{y,d,t,i}_{\mt{WT}} + P^{y,d,t,i}_{\mt{PV}} 
            \\ & \hspace{2em} + 
            P^{y,d,t,i}_{\mt{ESD}} - P^{y,d,t,i}_{\mt{ESC}} + 
            P^{y,d,t,i}_{\mt{BUY}} - P^{y,d,t,i}_{\mt{SELL}} 
        \end{aligned}
    \right] \label{st-p-a1}
    \\
    & \sum_{i \in \mathcal{N}_{i}} \left(P^{y,d,t,i}_{\mt{GEN}} -P^{y,d,t,i}_{\mt{DC}} -P^{y,d,t,i}_{\mt{LOAD}}\right) = 0 \label{st-p-a2}
    \\
    &P_{\mt{TL}}^{y,d,t,l} = \sum_{i \in \mathcal{N}_{i}} \Gamma_{l,i} (P^{y,d,t,i}_{\mt{GEN}} - P^{y,d,t,i}_{\mt{DC}} - P^{y,d,t,i}_{\mt{LOAD}} ) \label{st-p-a3}
    \\
    & -\mt{IC}_{\mt{TL}}^{y,l} \leq P_{\mt{TL}}^{y,d,t,l} \leq \mt{IC}_{\mt{TL}}^{y,l}, ~\forall l \in \mathcal{N}_{l} \label{st-p-a4}
    \\
    &0 \leq P^{y,d,t,i}_{\mt{BUY}},~P^{y,d,t,i}_{\mt{SELL}} \leq P^{\max}_{\mt{EX}} \label{st-p-a5}
\end{align}  
\label{pjm-opr-set-1}
\end{subequations}

The thermal unit capacity constraints are given in \eqref{pjm-opr-set-2}. Specifically, constraints \eqref{st-p-b1}-\eqref{st-p-b2} specify the capacity limitations for NG and CG units. Constraints \eqref{st-p-b3}-\eqref{st-p-b4} restrict the ramping capabilities accordingly. Equation \eqref{st-p-b5} ensures the initial output of the day equals the final output of the day.

\vspace{-1em}
\begin{subequations}
\begin{align}
    \beta_{\mt{NG}}^{\min} \mt{IC}_{\mt{NG}}^{y,i} &\leq P_{\mt{NG}}^{y,d,t,i} \leq \beta_{\mt{NG}}^{\max} \mt{IC}_{\mt{NG}}^{y,i} \label{st-p-b1}
    \\
    \beta_{\mt{CG}}^{\min} \mt{IC}_{\mt{CG}}^{y,i} &\leq P_{\mt{CG}}^{y,d,t,i} \leq \beta_{\mt{CG}}^{\max} \mt{IC}_{\mt{CG}}^{y,i}
    \label{st-p-b2}
    \\
    -R_{\mt{NG}}^{\mt{down}} \mt{IC}_{\mt{NG}}^{y,i} &\leq P_{\mt{NG}}^{y,d,t,i} - P_{\mt{NG}}^{y,d,t-1,i} \leq R_{\mt{NG}}^{\mt{up}} \mt{IC}_{\mt{NG}}^{y,i} \label{st-p-b3}
    \\
    -R_{\mt{CG}}^{\mt{down}} \mt{IC}_{\mt{CG}}^{y,i} &\leq P_{\mt{CG}}^{y,d,t,i} - P_{\mt{CG}}^{y,d,t-1,i} \leq R_{\mt{CG}}^{\mt{up}} \mt{IC}_{\mt{CG}}^{y,i} \label{st-p-b4}
    \\
    P_{\mt{NG}}^{y,d,0,i} &= P_{\mt{NG}}^{y,d,T,i},~P_{\mt{CG}}^{y,d,0,i} = P_{\mt{CG}}^{y,d,T,i} \label{st-p-b5}
\end{align}
\label{pjm-opr-set-2}
\end{subequations}
\vspace{-1em}

The operational constraints of renewable and storage units are formulated in \eqref{pjm-opr-set-3}. Specifically, constraints \eqref{st-p-c1} enforce the RES output limit. Equation \eqref{st-p-c2} describes the energy balance of the ES system. Constraints \eqref{st-p-c3} limit the energy capacity. Constraints \eqref{st-p-c4} limit the charging and discharging power. Equations \eqref{st-p-c5} enforce the starting and ending energy levels of energy storage systems to be equal.

\vspace{-1em}
\begin{subequations}
\begin{align}
    &0 \leq \left[P_{\mt{WT}}^{y,d,t,i},~P_{\mt{PV}}^{y,d,t,i} \right] \leq \left[ P_{\mt{WT,cap}}^{y,d,t,i},~P_{\mt{PV,cap}}^{y,d,t,i} \right] \label{st-p-c1}
    \\
    &E_{\mt{ES}}^{y,d,t,i} = E_{\mt{ES}}^{y,d,t-1,i} + (P_{\mt{ESC}}^{y,d,t,i}\eta_{\mt{esc}} - P_{\mt{ESD}}^{y,d,t,i}/\eta_{\mt{esd}}) \Delta t \label{st-p-c2}
    \\
    &\beta_{\mt{ES}}^{\min} \mt{IC}_{\mt{ES}}^{y,i} \leq E_{\mt{ES}}^{y,d,t,i} \leq \beta_{\mt{ES}}^{\max} \mt{IC}_{\mt{ES}}^{y,i} 
    \label{st-p-c3}
    \\
    &\left[P_{\mt{ESC}}^{y,d,t,i},P_{\mt{ESD}}^{y,d,t,i}\right] \begin{cases}
        \geq \left[ \beta_{\mt{ESC}}^{\min},\beta_{\mt{ESD}}^{\min} \right] \mt{IC}_{\mt{ES}}^{y,i}
        \\
        \leq \left[ \beta_{\mt{ESC}}^{\max},\beta_{\mt{ESD}}^{\max} \right] \mt{IC}_{\mt{ES}}^{y,i}
    \end{cases} \label{st-p-c4}
    \\
    &E_{\mt{ES}}^{y,d,0,i} = E_{\mt{ES}}^{y,d,T,i} \label{st-p-c5}
\end{align}
\label{pjm-opr-set-3}
\end{subequations}
\vspace{-1em}

\subsection{Operational constraints for data center} 
The constraints in formulation \eqref{dc-opr-set} describe data center operations ($\forall y \in \mathcal{N}_{y}, d \in \mathcal{N}_{d}, t \in \mathcal{N}_{t}, i \in \mathcal{N}_{i}, m \in \mathcal{N}_{m}$). Specifically, equation \eqref{st-dc-1} describes nodal computing workload balance, where $Y_{\mt{DC}}^{y,d,t,i}$ denotes computing service provided by the data center at node $i$, $Y_{\mt{TC}}^{y,d,t,ij}$ represents computing workload transferred from node $j$ to node $i$, and $Y_{\mt{LOAD}}^{y,d,t,i}$ denotes nodal computing workload demand. Constraints \eqref{st-dc-2} represent computing workload transfer limitations. Equation \eqref{st-dc-3} calculates computing workload provision at node $i$ based on active rack servers $X_{\mt{SVR},m}$. Equation \eqref{st-dc-4} calculates data center power consumption at node $i$, considering power utilization efficiency $\mt{PUE}_{m}$ and locational-seasonal cooling efficiency factor $\phi^{\mt{temp}}_{y,d,t,i}$. Constraint \eqref{st-dc-5} ensures active rack servers do not exceed installed servers.

\vspace{-1em}
\begin{subequations}
\begin{align}
    &Y_{\mt{DC}}^{y,d,t,i} + \sum_{\forall j, j \neq i} Y_{\mt{TC}}^{y,d,t,ij} - Y_{\mt{LOAD}}^{y,d,t,i} = 0 \label{st-dc-1}
    \\
    &-Z_{ij}\bar{Y}_{ij}^{y} - Y_{ij}^{0}\leq Y_{\mt{TC}}^{y,d,t,ij} \leq Z_{ij}\bar{Y}_{ij}^{y} + Y_{ij}^{0}\label{st-dc-2}
    \\
    &Y_{\mt{DC}}^{y,d,t,i} = \sum_{m \in \mathcal{N}_{m}} \psi_{\mt{SVR},m} X^{y,d,t,i}_{\mt{SVR},m} \Delta t\label{st-dc-3}
    \\
    &P_{\mt{DC}}^{y,d,t,i} = \sum_{m \in \mathcal{N}_{m}} \phi_{y,d,t,i}^{\mt{temp}} \hspace{0.5mm} \mt{PUE}_{m} X^{y,d,t,i}_{\mt{SVR},m} P^{\mt{rate}}_{\mt{SVR,m}} \label{st-dc-4}
    \\
    &0 \leq X^{y,d,t,i}_{\mt{SVR},m} \leq \mt{IN}_{\mt{SVR},m}^{y,i} \label{st-dc-5}
\end{align}
\label{dc-opr-set}
\end{subequations}

\vspace{-2em}
\section{Model reformulation and uncertainty capture}

In this section, we employ several solution techniques including single-objective reformulation, uncertainty sampling, and Benders decomposition. These techniques collectively transform the joint planning model into a tractable form.

\vspace{-0.5em}
\subsection{Planning model reformulation}
The joint planning model simultaneously considers both economic and environmental objectives. Given the complexity of multi-objective optimization problems, we employ the weighted sum method \cite{sum-weighted} to convert the multi-objective formulation into a single-objective form, as expressed in \eqref{single-obj}. In this formulation, $\hbar$ represents the weighting factor that reflects the decision maker's preference for carbon emission reduction and serves as a quantifiable parameter interpretable as the carbon emission price. Specifically, we adopt carbon emission auction price data from the Regional Greenhouse Gas Initiative (RGGI), using an initial price of \$22/short ton with a 2.3\% annual increase, as projected in the exploratory policy scenario \cite{PJM_Carbon_Price}.

\vspace{-0.5em}
\begin{equation}
\begin{aligned}
    J &= \min_{\mathbf{z} \in \mathbb{Z},~\mathbf{x} \in \mathbb{X}} \left[\mt{TC}(\mathbf{z,x}) +  \hbar \cdot \mt{TE}(\mathbf{z,x}) \right]
    \\
    \text{s.t.}& \hspace{1.5cm}\text{Constraints } \eqref{obj-eco}-\eqref{dc-opr-set}
\end{aligned}     \label{single-obj}
\end{equation}
\vspace{-1em}

\begin{figure}[!tb]
    \vspace{1.5em}
    \centering
    \begin{subfigure}[b]{0.48\linewidth}
        \centering
        \includegraphics[width=\linewidth]{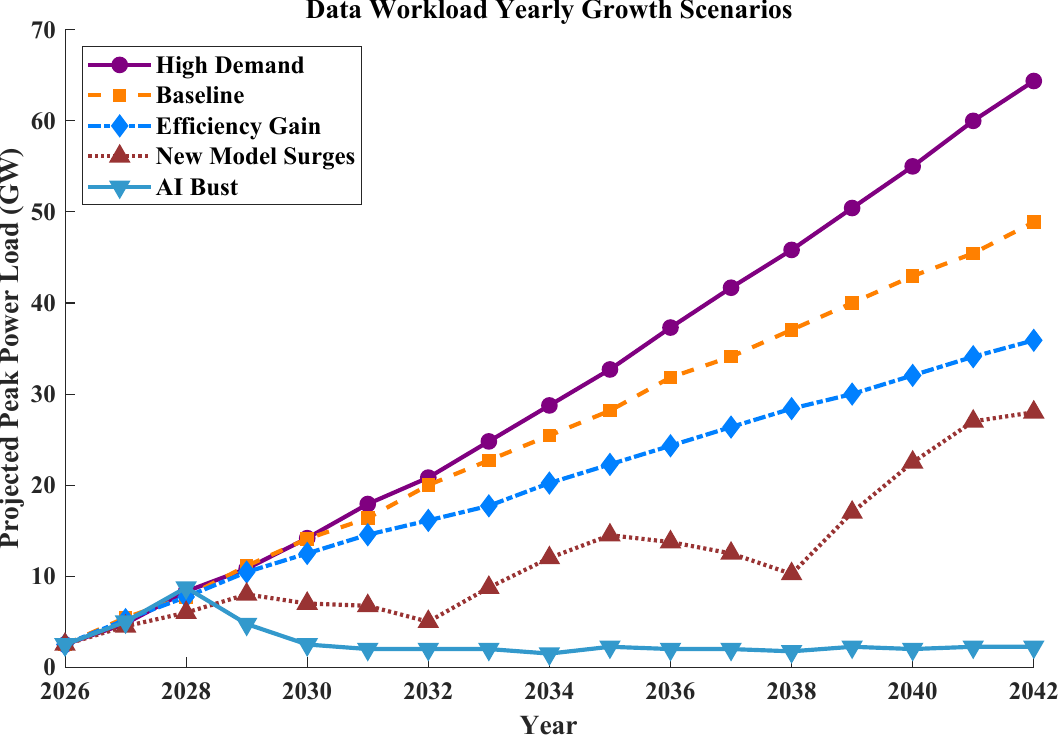}
        \caption{Yearly growth.}
        \label{fig-dc-annual}
    \end{subfigure}%
    \hfill
    \begin{subfigure}[b]{0.48\linewidth}
        \centering
        \includegraphics[width=\linewidth]{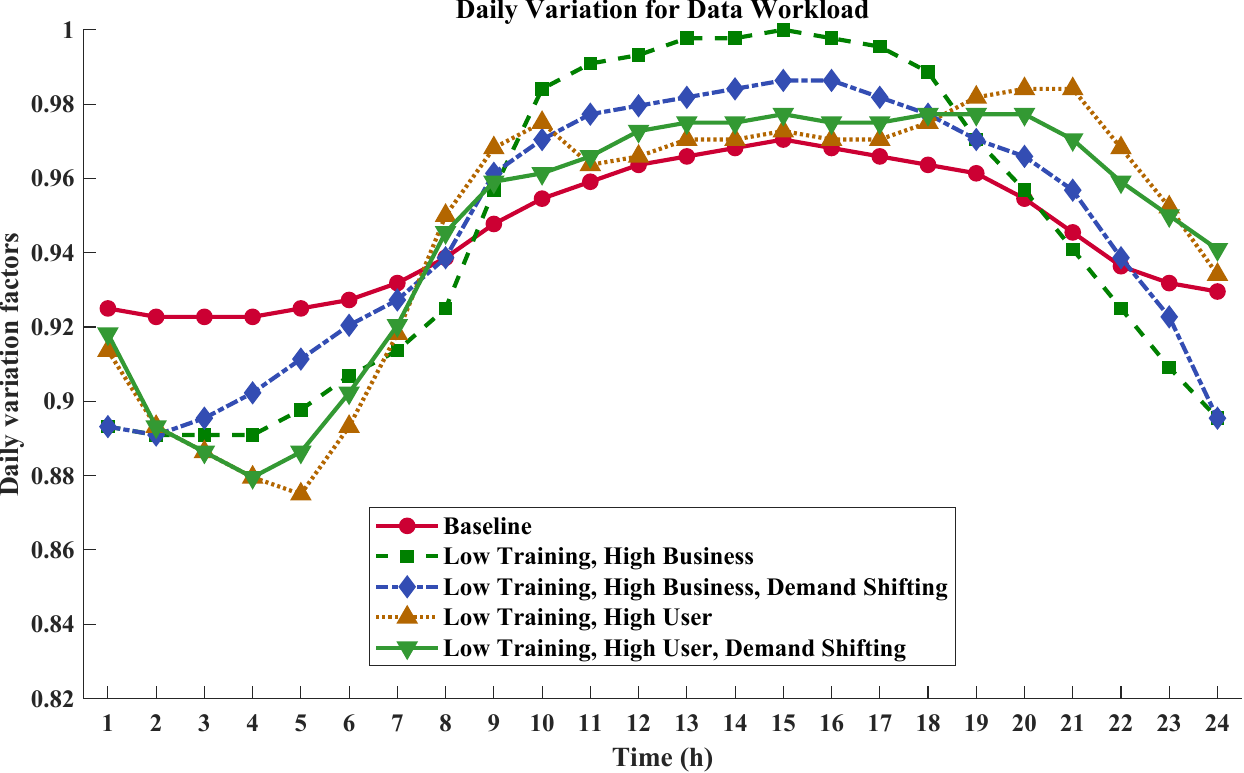}
        \caption{Daily variations.}
        \label{fig-dc-day}
    \end{subfigure}
    \vspace{-0.5em}
    \caption{AI demand variation patterns across temporal scales: (a) yearly growth trends and (b) daily fluctuations.}
    \label{fig-dc-combined}
    \vspace{-1em}
\end{figure}

\subsection{Long-Term and Short-Term Uncertainty Modeling} \label{scn-method}

We consider two classes of system uncertainties: long-term (e.g., annual growth in power demand and computing workload) and short-term (e.g., intra-day variations in both demands and the variability of solar and wind). 
Let $\zeta_{s}$ denote the stochastic scenario, incorporated into the planning model via \eqref{un-1}–\eqref{un-3} ($\forall y \in \mathcal{N}_{y}, d \in \mathcal{N}_{d}, t \in \mathcal{N}_{t}, \forall i \in \mathcal{N}_{i}$). Equation \eqref{un-1} specifies total power demand as the product of an annual increase factor $\lambda^{\mt{load}}_{y}$, a daily variation factor $\delta^{\mt{load}}_{d,t}$, and the base peak load $P^{\mt{peak}}_{\mt{LOAD}}$. The annual factor captures long-term uncertainty and is set to an average of 1\% per year, while the daily factor reflects short-term fluctuations obtained from historical data using the same sampling approach adopted later for RES daily variation.
\begin{align}
    &P^{y,d,t,i}_{\mt{LOAD}} = \varphi_{i}^{\mt{load}} \lambda^{\mt{load}}_{y} (\zeta_{s}) \hspace{0.5mm} \delta^{\mt{load}}_{d,t}(\zeta_{s}) \hspace{0.5mm} P^{peak}_{\mt{LOAD}} \label{un-1}
    \\
    &Y_{\mt{LOAD}}^{y,d,t,i} = \varphi_{i}^{\mt{data}} \lambda^{\mt{data}}_{y} (\zeta_{s})\delta^{\mt{data}}_{d,t}(\zeta_{s}) Y_{\mt{LOAD}}^{\mt{peak}} \label{un-2}
    \\
    &\left[P_{\mt{WT,cap}}^{y,d,t,i},~ P_{\mt{PV,cap}}^{y,d,t,i}\right] = 
    \left[
    \delta^{\mt{wt}}_{d,t}(\zeta_{s})\mt{IC}^{y,i}_{\mt{WT}},~\delta^{\mt{pv}}_{d,t}(\zeta_{s}) \mt{IC}^{y,i}_{\mt{PV}}
    \right] \label{un-3}
\end{align}

Equation \eqref{un-2} models the system’s computing workload demand—assumed AI-dominated—estimated by dividing the total power consumption of data centers in PJM by the average energy per request. The annual increase factor $\lambda^{\mt{dc}}_{y}$ spans five instances (high demand, baseline, efficiency gains, new model surges, AI bust), and the resulting yearly peak computing workload $\lambda^{\mt{dc}}_{y} Y_{\mt{LOAD}^{\mt{peak}}}$ is shown in Fig.~\ref{fig-dc-annual} \cite{AI-demand}. We adopt the general metric $\text{GPU}\cdot\text{h}$, benchmarking on NVIDIA V100 and assuming $1~\text{GPU}\cdot\text{h}=10^{4}$ requests; daily computing workload follows five representative instances in Fig.~\ref{fig-dc-day} \cite{AI-demand}. Equation \eqref{un-3} yields available RES output by scaling installed capacity with time-varying factors $\delta_{d,t}^{\mt{wt/pv}}$. To capture demand–supply uncertainties, we employ a hierarchical sampling strategy \cite{OPR-12}: (i) extract seasonal patterns and short-term fluctuations from historical data or distributions (e.g., wind via a Weibull distribution \cite{sample-1}, electric demand via hourly normal variations \cite{sample-2}); (ii) generate stochastic scenarios using Latin Hypercube Sampling that preserves cross-variable correlations \cite{sample-3}; and (iii) reduce the scenario set using $K$-means clustering to balance computational efficiency and accuracy.

In this study, seasonal variations are applied only to the power system load and renewable generation, as both electricity demand and renewable outputs exhibit systematic seasonal cycles that significantly affect long-term system planning. By contrast, the data center workload is assumed to be season-independent, reflecting empirical evidence that digital service demand remains stable throughout the year with limited correlation to climatic or seasonal factors.

\section{Benders decomposition method}
Direct solution is rendered intractable by the planning model’s two-stage stochastic structure and large temporal-scenario scale. Benders decomposition naturally separates planning and recourse decisions, reduces dimensionality, and exploits scenario block structure. As a proven method in large-scale optimization, it ensures computational efficiency and robust convergence for the proposed model.

\subsection{Compact formulation} 
The compact model is presented in \eqref{model-compact}, where $z_{y}$ denotes the planning decision variables for year $y$, and $x_{y,d,s}$ represents the operational decision variables for year $y$, day $d$, and scenario $s$. The parameter $\pi_{s}$ in \eqref{cpt-0} specifies the probability of scenario $s$. Constraint \eqref{cpt-1} corresponds to the planning constraints, where equalities are reformulated as pairs of inequalities for conciseness. Constraints \eqref{cpt-2} and \eqref{cpt-3} characterize the operational constraints for each day of each year.
\begin{subequations}
\begin{align}
    &J = \min \sum_{y \in \mathcal{N}_{y}} \left( c^{T}_{y} z_{y} + \sum_{d \in \mathcal{N}_{d}} \sum_{s \in \mathcal{N}_{s}} \pi_{s} q^{T} x_{y,d,s} \right) \label{cpt-0}
    \\
    \text{s.t. }& A_{y} z_{y} \leq b_{y},~\forall y \label{cpt-1}
    \\
    &E_{y,d}~x_{y,d,s} = f_{y,d,s}, ~\forall y,d, \forall s \in \mathcal{N}_{s} \label{cpt-2}
    \\
    & F_{y,d,s}~z_{y} + G_{y,d}~x_{y,d,s} \leq h_{y,d,s},~ \forall y,d, \forall s \in \mathcal{N}_{s} \label{cpt-3}
    \\
    & z_{y} \in \mathbb{Z}_{+}, ~ x_{y,d,s} \in \mathbb{R},
\end{align} \label{model-compact}
\end{subequations}
\vspace{-1em}

\vspace{-1em}
\subsection{Feasibility-check subproblem} 
We define a subproblem ($\mt{FP}_{y,d,s}$) to assess the feasibility of a planning-stage solution $\hat{z}_{y}$ with respect to the operational constraints under each daily stochastic scenario. The formulation of $\mt{FP}_{y,d,s}$ is provided in \eqref{fc-problem}. Specifically, objective function \eqref{fp-1} minimizes the sum of all non-negative slack variables. Constraints \eqref{fp-2} and \eqref{fp-3} impose the operational requirements while incorporating slack variables $\delta_{y,d,s}^{+}$, $\delta_{y,d,s}^{-}$, and $\varepsilon_{y,d,s}$. Equation \eqref{fp-4} enforces the planning decisions $z_{y}$ to match the obtained planning-stage solution, with $\mu_{y,d,s}$ denoting the associated dual variables. 
\begin{subequations}
\begin{align}
    \mt{FP}_{y,d,s}:&~ \Tilde{\Upsilon}_{y,d,s} = \min \left[ \mathbf{1}^{T} (\delta_{y,d,s}^{+} + \delta_{y,d,s}^{-}) + \mathbf{1}^{T} \varepsilon_{y,d,s} \right] \!\! \label{fp-1}
    \\
    \text{s.t. }
    &E_{y,d}~x_{y,d,s} = f_{y,d,s} + \delta_{y,d,s}^{+} - \delta_{y,d,s}^{-} \label{fp-2}
    \\
    & F_{y,d,s}~z_{y} + G_{y,d}~x_{y,d,s} \leq h_{y,d,s} + \varepsilon_{y,d,s} \label{fp-3}
    \\
    & z_{y} = \hat{z}_{y} ~~~ (\text{dual vars: }\lambda_{y,d,s}) \label{fp-4}
    \\
    & \delta_{y,d,s}^{+}, \delta_{y,d,s}^{-}, \varepsilon_{y,d,s} \geq 0 \label{fp-5}
\end{align}
\label{fc-problem}
\end{subequations}
\vspace{-1em}

\vspace{-1em}
\subsection{Optimality subproblem} 
The optimality subproblem $\mt{OP}_{y,d,s}$ is defined in \eqref{opt-formulation}. It is solved only when the feasibility-check problem attains a zero objective, indicating that the planning-stage solution satisfies all operational constraints. Specifically, objective \eqref{op-1} minimizes the operational recourse cost. Constraints \eqref{op-2}–\eqref{op-3} define the operational constraints under a planning decision $\hat{z}_{y}$. Equation \eqref{op-4} follows the proposed technique to capture the dual variables of $\mt{OP}_{y,d,s}$ with respect to the planning-stage solution.

\vspace{-1.5em}
\begin{subequations}
\begin{align}
    \mt{OP}_{y,d,s}:&~ \Tilde{\Phi}_{y,d,s} = \min q_{y}^{T} x_{y,d,s} \label{op-1}
    \\
    \text{s.t. }
    &E_{y,d}~x_{y,d,s} = f_{y,d,s} \label{op-2}
    \\
    & F_{y,d,s}~z_{y} + G_{y,d}~x_{y,d,s} \leq h_{y,d,s} \label{op-3}
    \\
    & z_{y} = \hat{z}_{y} ~~~ (\text{dual vars: } \mu_{y,d,s}) \label{op-4}
\end{align}
\label{opt-formulation}
\end{subequations}
\vspace{-1.5em}

\vspace{-1em}
\subsection{Enhanced algorithm}

\begin{algorithm}[!t] 
\caption{Benders decomposition method}
\label{alg-Benders}
\begin{algorithmic}
\State \textbf{Step 1.} Set $\mt{LB} = -\infty$, $\mt{UB} = \infty$, and $l = 0$.
\State \textbf{Step 2.} Solve the master problem.
\begin{itemize}
    \item If $\mt{MP}_{l}$ is infeasible, then terminate the algorithm and report the infeasibility.
    \item Otherwise, let $l \gets l + 1$. Get solution $z^{l,*}_y,\forall y $ and the optimal value $O^*$. Then, update $\mt{LB} = O^*$. 
\end{itemize}
\State \textbf{Step 3.}  Solve the subproblems $\mt{FP}$ and $\mt{OP}$ for each scenario $s \in \mathcal{N}_{s}$ as below.
\For{$\forall y \in \mathcal{N}_{y}$}
        \For{$\forall d \in \mathcal{N}_{d}$}
            \State Solve the feasibility-check problem $\mt{FP}_{y,d,s}$.
            \begin{itemize}[leftmargin=1.5cm]
                \item If $\Tilde{\Upsilon}_{y,d,s}^{l,*} = 0$, then set $\lambda_{y,d,s}^{l,*} = 0$, and solve the $\mt{OP}_{y,d,s}$ to derive optimal solution $\Tilde{\Phi}_{y,d,s}^{l,*}$, $\mu_{y,d,s}^{l,*}$.
                \item Otherwise, derive the optimal solution $\Tilde{\Upsilon}_{y,d,s}^{l,*} $ and $\lambda_{y,d,s}^{l,*}$. Set $\Tilde{\Phi}_{y,d,s}^{l,*} = 0$, $\mu_{y,d,s}^{l,*} = 0$.
            \end{itemize}
        \EndFor
\EndFor
\State \textbf{Step 4.}  Update $\mt{UB} = \min\{\mt{UB}, \mt{UB}^{l,*}$\}, where, 
\vspace{-1em}
\begin{equation}
    \hspace{0.6cm} \mt{UB}^{l,*} = \sum_{y \in \mathcal{N}_{y}} \left(c^{T}_{y} z_{y}^{l,*} + \sum_{d \in \mathcal{N}_{d}} \sum_{s \in \mathcal{N}_{s}} \pi_{s} \Tilde{\Phi}_{y,d,s}^{l,*}  \right) 
\end{equation}
\vspace{-1em}
\State \textbf{Step 5.} Check the tolerance and generate Benders cuts.
\begin{itemize}
    \item If $|\mt{UB}-{LB}|\leq $ tolerance, then terminate and report the optimal solution.
    \item Otherwise, generate and add Benders cuts in \eqref{benders-cut-1} and \eqref{benders-cut-2}. Go to the \textbf{Step 2.}
\end{itemize}
\end{algorithmic}
\end{algorithm}

With the above defined subproblems, the Benders decomposition method is presented in Algorithm \ref{alg-Benders}. The master problem (MP) is defined in \eqref{mp-formulation}.

\textit{\textbf{Remark:}} Several enhancement techniques for the Benders decomposition algorithm are implemented:  
\begin{itemize}
    \item The lower-level subproblems in Algorithm \ref{alg-Benders} are decomposed based on scenarios, which provides a well-structured framework to leverage the advantages of parallel computing, thereby improving computational efficiency.
    \item In the current framework in \textbf{Step 3}, subproblems are divided by individual uncertain scenarios. In practical coding, grouping several scenarios into a single subproblem can yield better computational efficiency. The number of scenarios included in each subproblem is determined heuristically. In this study, grouping four scenarios per subproblem yields the best performance.
    \item The master problem can include an initial operating scenario to expedite the feasibility check of the subproblems. In this study, we heuristically incorporate the scenario corresponding to the peak load into the master problem as the initial scenario, significantly enhancing the overall computational efficiency.
\end{itemize}

\vspace{-2em}
\begin{subequations}
\begin{align}
    \mt{MP}_{l}:~& O^{*} = \min \sum_{y \in \mathcal{N}_{y}} \left(c^{T}_{y} z_{y} + \sum_{d \in \mathcal{N}_{d}} \sum_{s \in \mathcal{N}_{s}} \pi_{s} \Tilde{\Phi}_{y,d,s}  \right) \!\!
    \\
    \text{s.t. } & A_{y} z_{y} \leq b_{y},~\forall y 
    \\
    &\Tilde{\Phi}_{y,d,s} \geq \Tilde{\Phi}_{y,d,s}^{j,*} - \mu_{y,d,s}^{j,*}(z_{y} - z_{y}^{j,*}), \nonumber
    \\
    &\hspace{11em}\forall y,d,s,~1 \leq j \leq l \label{benders-cut-1}
    \\
    &0 \geq \Tilde{\Upsilon}_{y,d,s}^{j,*} - \lambda_{y,d,s}^{j,*}(z_{y} - z_{y}^{j,*}),~\forall y,d,s,~1 \leq j \leq l \label{benders-cut-2}
    \\
    & z_{y} \in \mathbb{Z}_{+}, \Tilde{\Phi}_{y,d,s} \in \mathbb{R}_{+}, ~~~~\forall y,d,s
\end{align}
\label{mp-formulation}
\end{subequations}
\vspace{-1em}

\vspace{-1em}
\section{Case study}

\subsection{System configuration}
\begin{figure}[!h]
    \centering
    \includegraphics[width= 3 in]{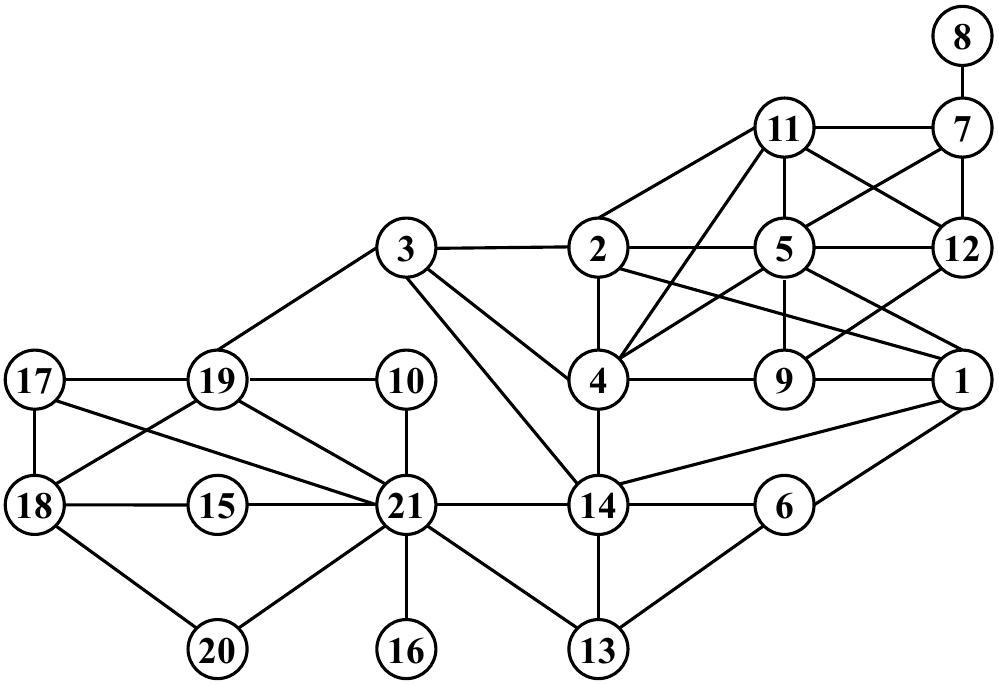}
    \vspace{-0.5em}
    \caption{Topology of the PJM 21-bus system.}
    \label{fig-topology}
\end{figure}

This paper evaluates the proposed joint planning model using the PJM system, one of the largest ISOs in the United States, which hosts about one-third of the nation’s data centers. PJM system information are obtained from the U.S. Energy Information Administration (EIA) \cite{EIA2022}. PJM transmission zones are defined in \cite{PJMMaps}, each modeled as a node, and transmission lines above 161 kV between zones are considered. The simplified system consists of 21 buses and 42 transmission corridors, the topology of the PJM-21-bus system is shown in \refig{fig-topology}, with the dataset available at \cite{github-pjm-case}. The annual peak PJM demand is 173 GW \cite{PJMDataMiner2}, the initial data center demand is 2.5 GW \cite{AI-demand, DOM-datacenter-demand}.

Computational Environment: The proposed model was implemented in MATLAB R2023a, leveraging the YALMIP toolbox for formulation and solved with Gurobi 11.0.0. All computations were conducted on a workstation featuring an Intel Core i9-9900X CPU and 64 GB of RAM. The associated GitHub repository provides the PJM case data in MATPOWER format.

Capital and operational cost data for generation resources are taken from \cite{capital-all, capital-line}. Embodied carbon emissions of generation resources are obtained from \cite{embCO2-coal, embCO2-line, embCO2-pv-wt, embCO2-sto}, while data for data center racks are derived from \cite{embCO2-dc, AI-demand}. The capital cost of servers is estimated from publicly available NVIDIA data. Other parameters are borrowed from \cite{OPR-8, OPR-12}. Operational scenarios for demand and RES are obtained from PJM API tools \cite{PJMDataMiner2}, and data center demands from \cite{AI-demand}. The collected data are summarized in Table \ref{tab-data-1}–\ref{tab-data-4}. 

\begin{table*}[!tbp]
    \centering
    \caption{Generation resources and energy storage devices}
    \vspace{-0.5em}
    \resizebox{\linewidth}{!}{%
    \begin{tabular}{lccccccc}
        \toprule
        \multicolumn{8}{c}{\textbf{Generation Resources}} \\
        \midrule
        Label & Capacity (MW) & Gen. Cost (\$/MWh) & Inv. (\$/MW) & O\&M (\$/MWh) & Emb CO$_2$ (lbs CO$_2$/MW) & Opr CO$_2$ (lbs CO$_2$/MWh) & Life Time (yr) \\
        \midrule
        CG  & 100  & 37.43 & 4,103,000  & 6.4  & 169,091  & 1400 & 30 \\
        NG  & 100  & 39.59 & 921,000   & 3.3  & 152,654  & 1200  & 30 \\
        PV  & 10  & /     & 1,302,000  & 2.8  & 1,485,900  & /   & 25 \\
        WT  & 10  & /     & 1,686,000  & 9.6  & 503,360   & /   & 25 \\
        \midrule
        \multicolumn{8}{c}{\textbf{Energy Storage Devices}} \\
        \midrule
        Label & Energy Rating (MWh) & Power Rating (MW) & Energy Inv. (\$/MWh) & Power Inv. (\$/MW) & O\&M (\$/MWh) & Emb CO$_2$ (lbs CO$_2$/MWh) & Life Time (yr) \\
        \midrule
        ES  & 1  & 0.35  & 436,000  & 1,245,000  & 4.2  & 240,000  & 18 \\
        \bottomrule
    \end{tabular}%
    \label{tab-data-1}
    }
\end{table*}

\begin{table}[!tbp]
    \centering
    \caption{Data center server module specifications}
    \vspace{-0.5em}
    \resizebox{\linewidth}{!}{%
    \begin{tabular}{lcccccccc}
        \toprule
        Label & Description & Unit  & Inv.  & Emb CO$_2$  & Power  & Capability  & PUE  & Life Time \\
              &            & (Servers/Rack) & (\$/Rack) & (lbs CO$_2$/Rack) & (kW/Rack) & (GPU$\cdot$h/Rack) &  & (yr) \\
        \midrule
        Fix Inv.    & Buildings \& Racks        & / & 850,000   & 150,000    & /  & / & / & 25 \\
        Type-I   & V100 \& Air Cooling   & 150  & 1,500,000  & 900,000  & 50   & 150   & 1.5 & 5 \\
        Type-II  & H100 \& Water Cooling  & 150  & 4,800,000  & 750,000  & 130  & 2,370  & 1.3 & 5 \\
        Type-III & H200 \& Liquid Cooling & 150  & 5,650,000  & 650,000  & 135  & 3,090  & 1.2 & 5 \\
        \bottomrule
    \end{tabular}%
    \label{tab-data-3}
    }
\end{table}

\begin{table}[!tbp]
    \raggedright
    \caption{Electricity and data transmission}
    \vspace{-0.5em}
    \resizebox{\linewidth}{!}{%
    \begin{tabular}{lcccc}
        \toprule
        \multicolumn{5}{c}{\textbf{Transmission Lines}} \\
        \midrule
        Label & Rating (MW) & Inv. (\$/km $\cdot$ MW) & Emb CO$_2$ (lbs CO$_2$/km $\cdot$ MW) & Life Time (yr) \\
        \midrule
        TL & 10 & 21,940 & 21,340 & 30 \\
        \midrule
        \multicolumn{5}{c}{\textbf{Data Transfer Facilities}} \\
        \midrule
        Label & Bandwidth (Gbps) & Inv. (\$/km) & Emb CO$_2$ (lbs CO$_2$/km) & Life Time (yr) \\
        \midrule
        TC & 50 & 60,000 & 17,637 & 30 \\
        \bottomrule
    \end{tabular}%
    \label{tab-data-2}
    }
\end{table}

\begin{table}[!tbp]
    \raggedright
    \caption{Economic and technical parameters}
    \vspace{-0.5em}
    \resizebox{\linewidth}{!}{%
    \begin{tabular}{llc}
        \toprule
        Parameter & Description & Value \\
        \midrule
        $\sigma$ & Discount rate & 0.04 \\
        $R^{0}$ & Residual factor at retirement & 0.15 \\
        \multirow{3}{*}{$\xi^{\mt{buy/sell}}$} & On-peak Electricity price (\$/MWh) & 156 \\
        & Mid-peak Electricity price (\$/MWh) & 115 \\
        & Off-peak Electricity price (\$/MWh) & 82 \\     
        $\xi^{\mt{curt}}_{\mt{PV/WT}}$ & Curtailment cost for PV and WT (\$/MWh) & 30 \\
        $\tau_{0}$ & Propagation delay per unit distance (ms/100 km) & 0.51 \\
        $T_{\mt{req}}$ & Maximum propagation delay (ms) & 10 \\
        $\lambda_{\mt{tech}}$ & Embodied carbon equivalent rate from server tech advancements & 0.95 \\
        $\gamma_{\mt{tech}}$ & Investment equivalent rate from server tech advancements & 0.85 \\
        $\hbar$ & Initial carbon price (\$/short ton)  & 22 \\
        \bottomrule
    \end{tabular} }
    \label{tab-data-4}
\end{table}

\begin{figure}[!h]
    \centering
    \includegraphics[width= 3.2 in]{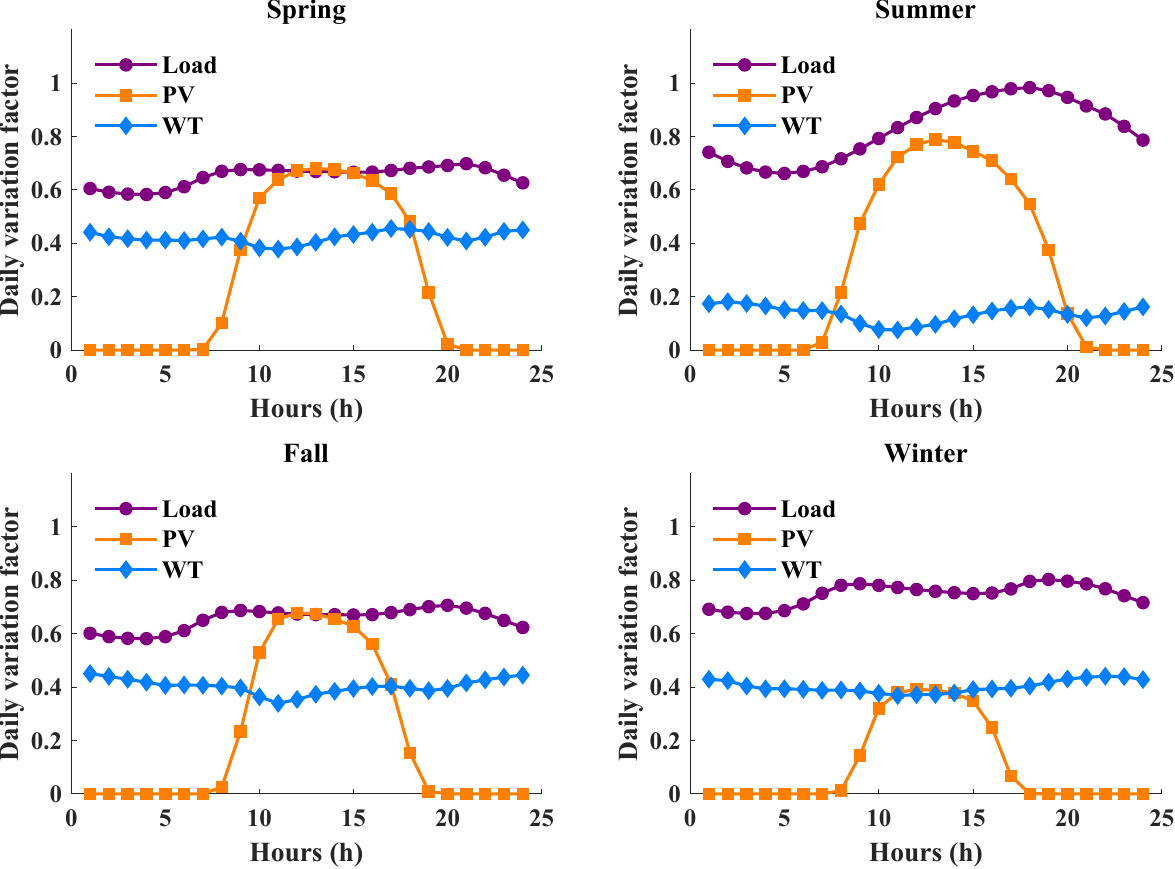}
    \vspace{-0.5em}
    \caption{Seasonal patterns for demand, solar generation, and wind generation.}
    \label{fig-scn}
\end{figure}

A 15-year planning horizon is considered, with four representative days per year to capture seasonal variation. In total, 50 operating scenarios are generated, accounting for long-term demand growth and short-term RES variability, using the method in Section \ref{scn-method}, which also provides daily and yearly patterns of data centers. Representative scenarios are illustrated in \refig{fig-scn}. 

In this study, each operational scenario encompasses the entire 15-year planning horizon. Within each scenario, every year is represented by 4 typical days, with each day further divided into 24 hourly time steps, resulting in a total of \( 24 \times 4 \times 15 = 1,440 \) time periods per scenario. To capture uncertainties in demand and renewable generation, a total of 50 such scenarios are considered, leading to a large-scale stochastic optimization problem that motivates the use of Benders decomposition introduced in later sections.

\subsection{Performance comparison}
This section evaluates the proposed joint planning method against a conventional non-joint planning approach, demonstrating the benefits of co-optimizing power systems and data centers in terms of both cost efficiency and decarbonization. We further assess the role of embodied carbon emissions in planning, underscoring their significance for long-term sustainability.

\subsubsection{Comparison with non-joint planning approach}
Conventional non-joint planning approaches are widely adopted in the literature \cite{R6,R3-XJTU,R7}, where data centers determine siting and capacity based on locational carbon intensity and electricity prices, while power system planning is conducted independently according to projected demand growth. In contrast, the joint planning employs a system-level co-optimization framework that co-optimize both infrastructures.

Table~\ref{non-joint-comparison} summarizes the comparison results. Both methods meet the same total demand and service levels to ensure fairness, and the evaluation is conducted using a one-year static planning instance based on the PJM system. Subsequent sections extend the analysis to the full 15-year dynamic horizon. As shown in Table~\ref{non-joint-comparison}, the joint planning framework markedly outperforms the non-joint approach, achieving a 12.6\% reduction in total investment cost, an 8.3\% reduction in operational cost, and a 5.6\% decrease in operational carbon emissions. In addition, the coordinated planning of power systems and data centers increases the renewable capacity share by more than 25\%, demonstrating its effectiveness in advancing decarbonization.

\begin{table}[!tbp]
\centering
\caption{Comparison of joint and non-joint planning approaches}
\vspace{-0.5em}
\label{non-joint-comparison}
\resizebox{\linewidth}{!}{%
\begin{tabular}{lcccc}
\toprule
\textbf{Metric \hspace{30mm}} & \textbf{\hspace{5mm} Unit \hspace{5mm}} & \textbf{\hspace{5mm} Non-Joint \hspace{5mm}} & \textbf{Joint (Proposed)} & \textbf{\hspace{3mm} Improvement (\%) \hspace{3mm}} \\
\midrule
Total Investment Cost & B\$/yr & 48.76 & \textbf{42.60} & \textbf{$-$12.63\%} \\
\quad $\triangleright$ Power System Investment & B\$/yr & 28.43 & \textbf{25.10} & \textbf{$-$11.72\%} \\
\quad $\triangleright$ Data Center Investment & B\$/yr & 20.33 & \textbf{17.50} & \textbf{$-$13.93\%} \\
Operational Cost & B\$/yr & 62.19 & \textbf{57.05} & \textbf{$-$8.25\%} \\
Operational Carbon Emissions & MtCO\textsubscript{2}/yr & 154.7 & \textbf{145.9} & \textbf{$-$5.63\%} \\
Embodied Carbon Emissions & MtCO\textsubscript{2}/yr & 85.1 & \textbf{72.6} & \textbf{$-$14.69\%} \\
Renewable Capacity Share & \% & 36.2\% & \textbf{45.4\%} & \textbf{$+$25.41\%} \\
\bottomrule
\end{tabular}
}
\end{table}

\begin{table}[!tbp]
\centering
\caption{Comparison of joint planning with and without embodied carbon consideration on the 15-year PJM case}
\vspace{-0.5em}
\label{joint-em-comparison}
\resizebox{\linewidth}{!}{%
\begin{tabular}{lccc}
\toprule
\textbf{Metric} & \textbf{Joint-OP (Only OP in Objective)} & \textbf{Joint-EM (OP \& EM)} & \textbf{Change (\%)} \\
\midrule
Total Investment Cost (B\$)                            & 967.6     & \textbf{1035.3} & \textbf{$+$7.0\%} \\
Operational Carbon Emissions (MtCO\textsubscript{2})   & 2929.2    & \textbf{2434.0} & \textbf{$-$16.9\%} \\
Embodied Carbon Emissions (MtCO\textsubscript{2})      & 364.9     & \textbf{326.6}  & \textbf{$-$10.5\%} \\
Renewable Generation Capacity (GW)                     & 229.6     & \textbf{288.2}  & \textbf{$+$25.5\%} \\
Renewable Share (\% of total capacity)                 & 47.6\%    & \textbf{59.6\%} & \textbf{$+$25.2\%} \\
Storage Deployment (GWh)                               & 68.8      & \textbf{101.2}  & \textbf{$+$47.1\%} \\
RES Curtailment (\% of RES gen)                        & 13.2\%    & \textbf{8.4\%}  & \textbf{$-$36.4\%} \\
\bottomrule
\end{tabular}
}
\end{table}

\subsubsection{Impact of embodied carbon consideration}
To assess the role of embodied carbon emissions in long-term planning, we compare two variants of the joint model: (i) \textbf{Joint-OP}, which considers only operational emissions; and (ii) \textbf{Joint-EM}, which incorporates both operational and embodied emissions. Both are evaluated over a 15-year planning horizon under the High-Demand PJM instance.

Results in Table~\ref{joint-em-comparison} show that accounting for embodied emissions increases total investment cost by 7.0\%, driven by the adoption of low-carbon technologies and infrastructure. However, this additional investment yields substantial benefits: a 16.9\% reduction in operational emissions and a 10.5\% reduction in embodied emissions. Moreover, embodied carbon-aware planning promotes higher deployment of renewable energy (up 25.5\%) and storage capacity (up 47.1\%), while reducing renewable curtailment by more than one-third. These findings underscore the importance of including embodied emissions in long-term planning to avoid suboptimal strategies and to support deep decarbonization.

\begin{figure*}[!tbp]
    \centering
    \includegraphics[width=0.9\linewidth]{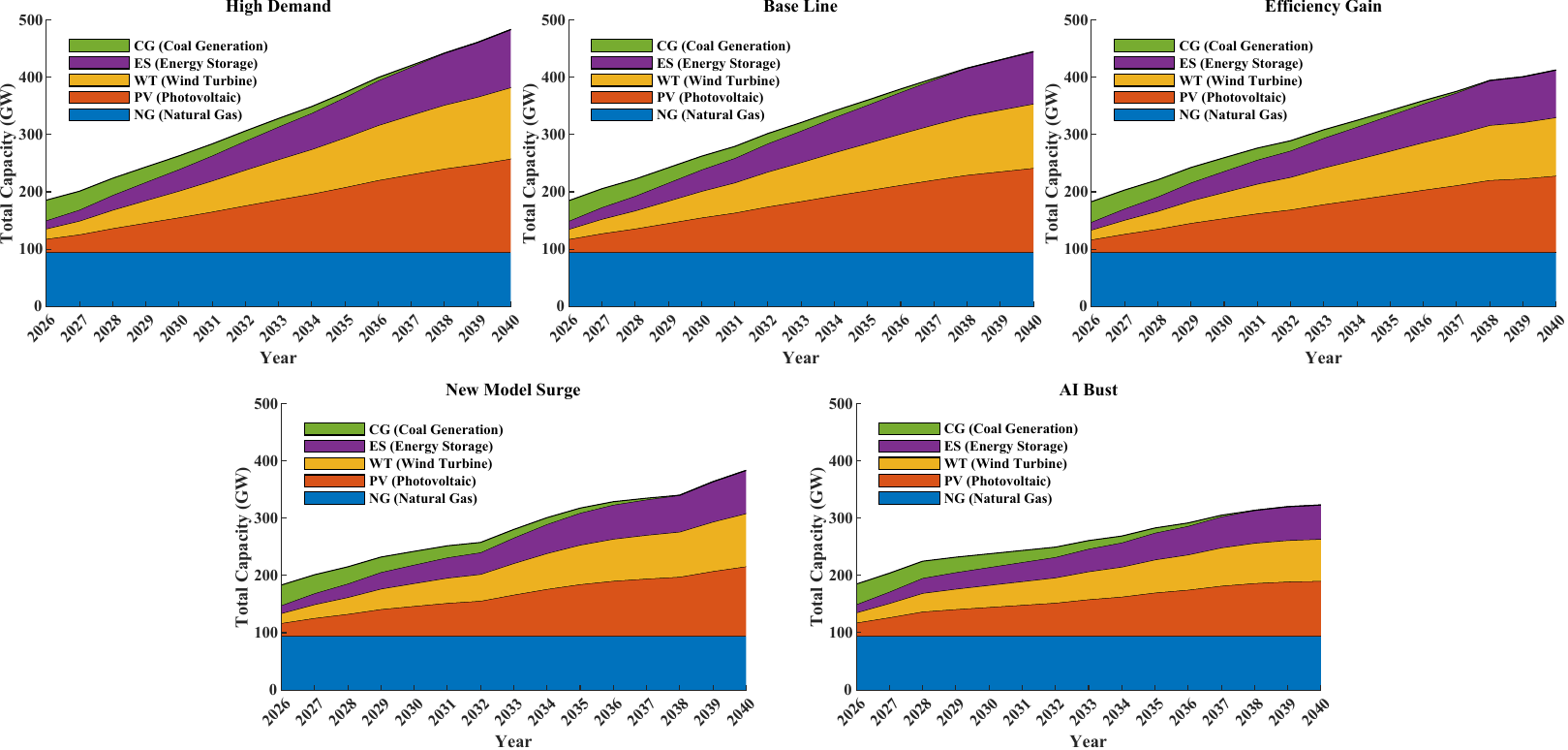}
    \caption{The long-term planning results for generation resources  under five computing workload demand conditions.}
    \label{fig-generation-resource-plan}
    \vspace{-1.5em}
\end{figure*}

\subsection{Analysis of capacity expansion results}
In this paper, we consider planning of NG, PV, WT, and ES units, along with CG retirements, while hydro and nuclear are treated as base-load in PJM. \refig{fig-generation-resource-plan} presents generation resource planning under five workload instances: High-Demand, Baseline, Efficiency-Gain, New Model Surge, and AI-Bust.

As shown in \refig{fig-generation-resource-plan}, RES capacity rises from 21.25 GW (10.6\% of PJM) in the base year to 389.4 GW (74.4\%) by 2040 in the High-Demand case, illustrating that most additional computing demand can be supported by renewables, particularly solar PV due to its correlation with workload profiles. Even under the AI-Bust instance, RES capacity reaches 228.9 GW (58.8\% of the High-Demand case). The Baseline and Efficiency-Gain instance yield 350 GW and 318 GW of RES by 2040 (89.9\% and 81.7\% of High-Demand), while the New Model Surge case delays expansion due to efficiency-driven demand troughs, resulting in 289 GW by 2040. Across all instances, the RES mix converges to PV:WT:ES $\approx$ 0.4:0.3:0.3, driven by PV’s cost advantage, policy mandates for WT, and ES’s role in mitigating variability. NG capacity remains unchanged since new additions conflict with decarbonization goals, while CG retirements consistently reach the annual 3 GW cap, underscoring the urgency of phasing out carbon-intensive units.

\begin{figure*}[!tbp]
    \centering
    \includegraphics[width= \linewidth]{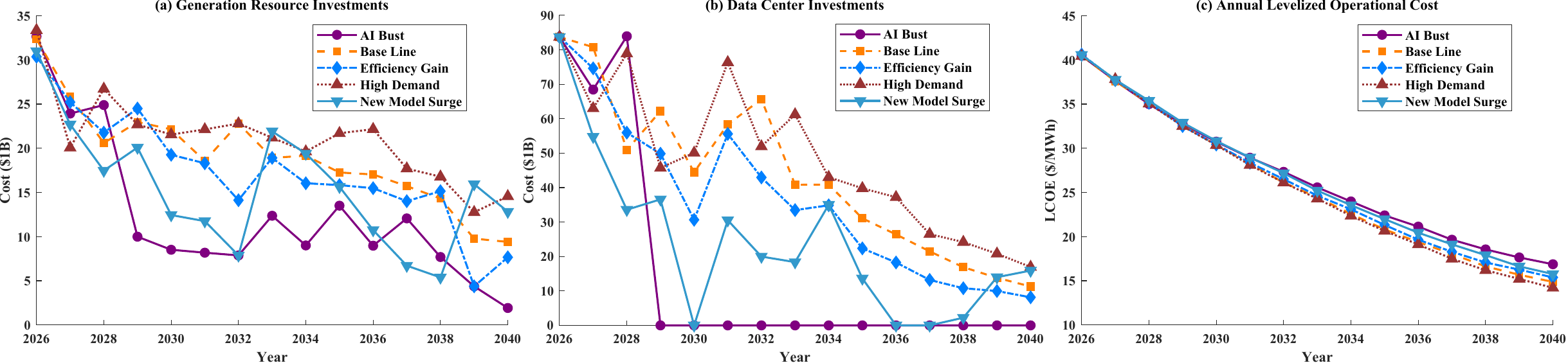}
    \caption{Cost analysis for joint planning, including (a) annual investments in power generation resources, (b) annual investments in data centers, and (c) annual operational costs, evaluated using the LCOE.}
    \label{fig-cost-info}
\end{figure*}

\begin{figure*}[!tbp]
    \centering
    \includegraphics[width= \linewidth]{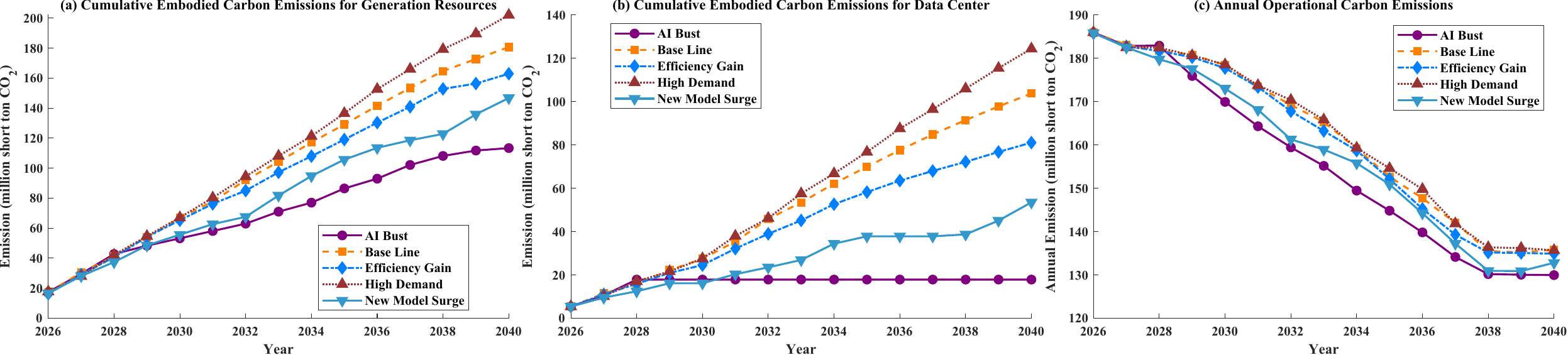}
    \caption{Carbon emission analysis for joint planning, including (a) cumulated embodied carbon emission for power generation resources, (b) cumulated embodied carbon emission for data centers, and (c) annual operational carbon emission of the whole system.}
    \label{fig-carbon-emission}
\end{figure*}

\begin{figure}[!tb]
    \centering
    \includegraphics[width= 0.6\linewidth]{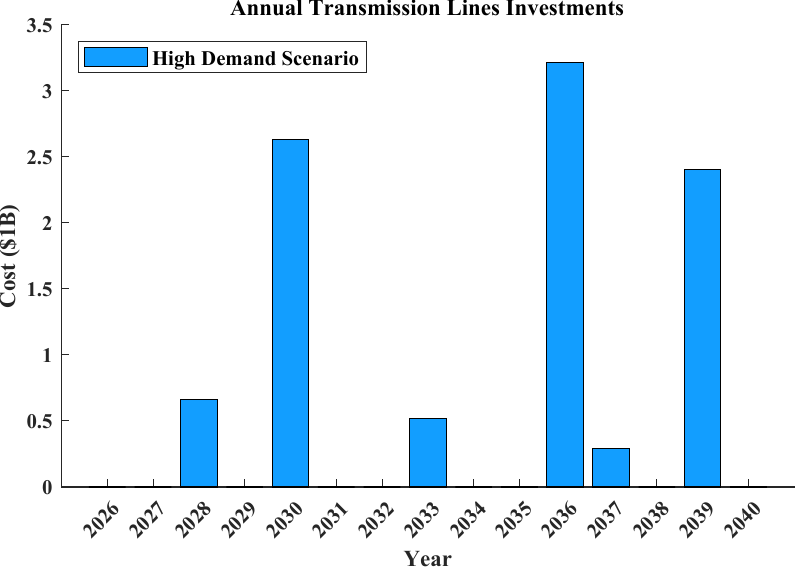}
    \vspace{-1em}
    \caption{Annual investments for power transmission lines under AI High-Demand instance.}
    \label{fig-transmission-cost}
\end{figure}

\subsection{Analysis of investment cost}

This section presents the annual investment and operational costs of generation resources and data centers, with results shown in \refig{fig-cost-info}. 

\refig{fig-cost-info}\textcolor{blue}{(a)} shows that annual generation investments, entirely from RES expansion—decline over time, from 33.36 billion USD in 2026 to 14.56 billion USD in 2040 under the High-Demand instance. The decline reflects both discounting effects and diminishing marginal benefits of RES deployment, as higher penetration requires increasing ES support. Total investment strongly correlates with AI demand, ranging from 389.4 billion USD in the High-Demand case to 185.9 billion USD in the AI-Bust case.

\refig{fig-cost-info}\textcolor{blue}{(b)} illustrates data center investment trends. Peak investments occur at the start, then decline with technology cost reductions (discount factor 0.85/year). Instance-specific effects are evident: New Model Surge shows sharp investment drops due to efficiency gains; AI-Bust falls to zero after 2029; while High-Demand sustains the highest investment levels. Baseline and Efficiency-Gain instance both show gradual deceleration after 2030.

\refig{fig-cost-info}\textcolor{blue}{(c)} depicts system-level LCOE, which falls from 40.4 \$/MWh in 2026 to 14.2 \$/MWh in 2040 (-65\%), driven by large-scale RES deployment. By 2040, operating costs diverge across instances: the High-Demand case achieves the lowest LCOE due to accelerated RES expansion, whereas AI-Bust has the highest due to limited RES penetration. These results highlight the critical role of AI-driven demand in shaping both investment and operational outcomes.

Transmission investment patterns are shown in \refig{fig-transmission-cost}. In the High-Demand case, congestion leads to discrete expansion stages: small investments in 2028 (0.66 billion) and 2033 (0.52 billion) are followed by major upgrades in 2030 (2.63 billion), 2036 (3.21 billion), and 2039 (2.4 billion). This stepwise pattern aligns with PJM practices and underscores the importance of joint planning across generation, transmission, and data center infrastructures.

\subsection{Analysis of carbon emissions}
This section analyzes the carbon emissions associated with the proposed long-term planning for PJM, including both embodied emissions from infrastructure construction and operational emissions from power generation. The results are summarized in \refig{fig-carbon-emission}.  

\refig{fig-carbon-emission}\textcolor{blue}{(a)} presents cumulative embodied carbon emissions from generation resource construction. In the early years, emissions are similar across instances, reaching about 50 million short tons of CO$_2$ by 2029. After 2030, trajectories diverge: by 2040, emissions reach 203.6 million in the High-Demand instance, compared with 178.2, 156.7, and 143.8 million in the Baseline, Efficiency-Gain, and New Model Surge cases, respectively. The AI-Bust instance records the lowest value, 108.6 million. The gap of 95 million between High-Demand and AI-Bust corresponds to 87.5\% of AI-Bust emissions, underscoring the dominant impact of AI-driven demand. Across instances, about 62.3\% of embodied emissions come from solar PV deployment.  

\refig{fig-carbon-emission}\textcolor{blue}{(b)} shows embodied emissions from data center construction. Under the High-Demand instance, emissions reach 128.6 million short tons by 2040, equal to 63.2\% of those from all generation resources in the same year. Baseline and Efficiency-Gain instances exhibit a steady increase, while the New Model Surge case follows a phased trajectory due to efficiency gains. In contrast, the AI-Bust instance stabilizes after 2028 as data center expansion halts. These results highlight the significant carbon footprint of GPU manufacturing in AI infrastructure.  

\refig{fig-carbon-emission}\textcolor{blue}{(c)} depicts operational carbon emissions, which decline overall due to large-scale RES integration. The evolution can be divided into three stages: a slow decrease during 2026–2028, a sharp reduction in 2029–2037 with clear instance differences, and a plateau in 2038–2040 as marginal benefits diminish. On average, operational emissions remain at 158.1 million short tons of CO$_2$ annually, reflecting the continued challenge of balancing AI-driven demand with decarbonization goals.

\begin{figure*}[!h]
    \centering
    \includegraphics[width= \linewidth]{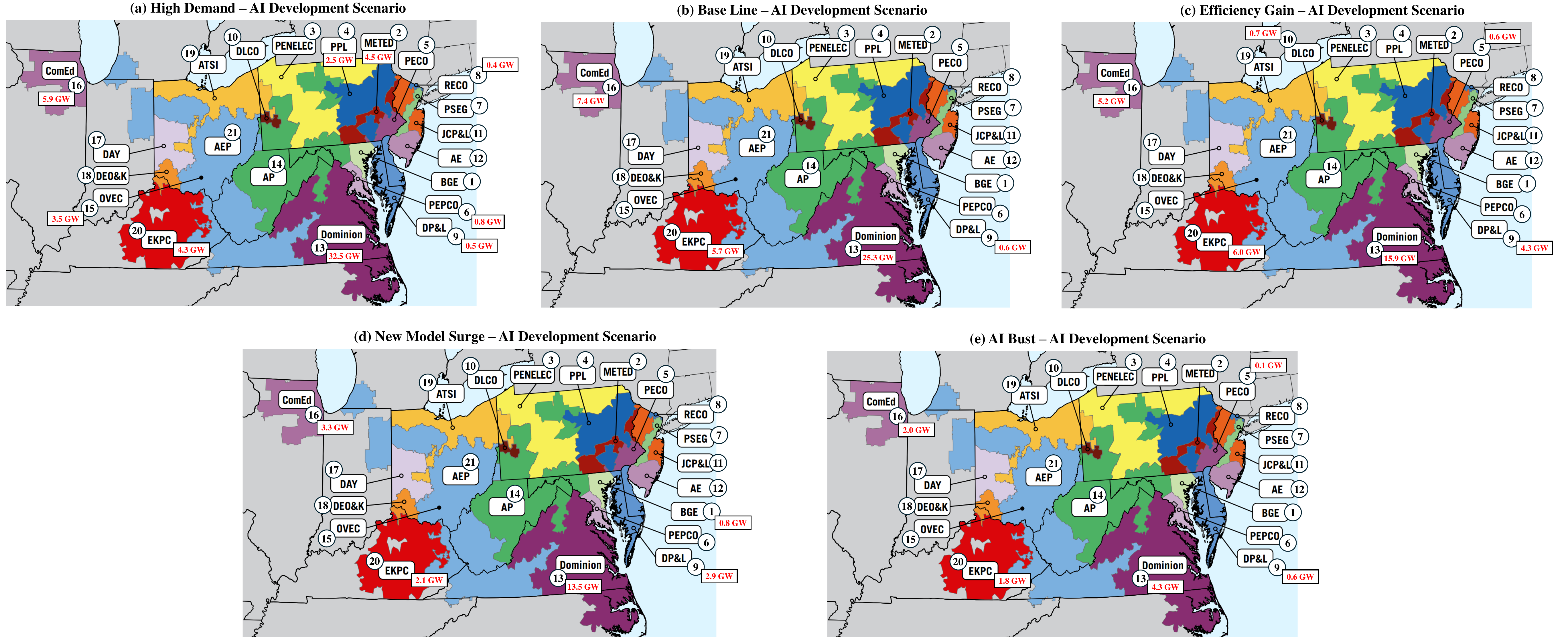}
    \vspace{-2em}
    \caption{The locational planning results for data centers in 2040 under five AI demand instances, the map is obtained from PJM \cite{PJMMaps}.}
    \label{fig-datacenter-plan}
\end{figure*}

\subsection{Data center construction}

This section analyzes the geographical distribution of data centers in 2040 under different AI development instances, as shown in \refig{fig-datacenter-plan}. Data center size is expressed in terms of equivalent electrical demand capacity, with locations assumed at the geometric centers of PJM transmission zones.

In the High-Demand instance \refig{fig-datacenter-plan}\textcolor{blue}{(a)}, data centers are distributed across nine zones, with Dominion (DOM) dominating at 32.5 GW, followed by ComEd (5.9 GW), METED (4.5 GW), EKPC (4.3 GW), and OVEC (3.5 GW). Other zones each host less than 3 GW. The concentration in DOM aligns with its existing role as a national data center hub near Richmond, Virginia, supporting the validity of the planning framework. The projected 32.5 GW can be regarded as an upper bound, compared with the current 2–5 GW load \cite{DOM-datacenter-demand}. Three factors explain DOM’s strategic advantage:  

\begin{enumerate}
    \item \textit{Abundant energy resources:} DOM has substantial RES and NG supply, enabling lower prices and emissions.  
    \item \textit{Robust communication infrastructure:} It serves as a hub for submarine cables and fiber networks, offering high-capacity, low-latency connectivity.  
    \item \textit{Well-integrated transmission network:} Strong interconnections mitigate congestion risks from large-scale data center deployment.  
\end{enumerate}

Across other instances \refig{fig-datacenter-plan}\textcolor{blue}{(b–e)}, data centers remain concentrated in DOM, ComEd, and EKPC, though with lower total capacities. For example, in the Baseline instance DOM hosts 25.3 GW, ComEd 7.4 GW, and EKPC 5.7 GW. The Efficiency-Gain, New Model Surge, and AI-Bust cases exhibit similar patterns, with DOM consistently holding the largest share.  

In summary, DOM emerges as the primary hub for future data center growth, while ComEd and EKPC are secondary hubs. These three zones should therefore be prioritized in long-term investment strategies.

\begin{figure*}[!tb]
    \centering
    \includegraphics[width=\linewidth]{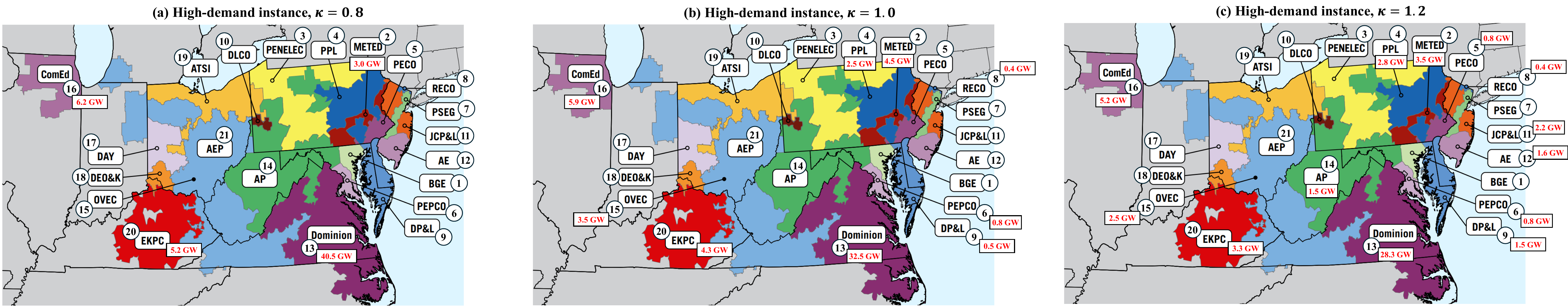}
    \vspace{-2em}
    \caption{Data center locational planning results under the high-demand instance with variations in the carbon cost factor.}
    \label{fig-carbon-cost}
\end{figure*}

\subsection{Sensitivity analysis on the carbon cost factor $\hbar$}

The carbon cost factor $\hbar$ serves as a critical parameter that connects data center planning decisions to system-wide carbon emissions. To evaluate the impact of $\hbar$, a sensitivity analysis is conducted using an index parameter $\kappa$, which scales the relative magnitude of $\hbar$. Three scenarios are examined, corresponding to $\kappa = 0.8$, $\kappa = 1.0$, and $\kappa = 1.2$. The analysis is conducted using the high-demand AI development case to evaluate the resulting variations in system performance and carbon outcomes.

Under the aforementioned settings, the proposed joint planning model is applied to obtain the corresponding optimal planning outcomes, as summarized in \refig{fig-carbon-cost}. As shown in the figure, notable variations in data center siting decisions emerge across the three scenarios. Specifically, when $\kappa = 0.8$, the majority of capacity is concentrated in cost-efficient yet carbon-intensive regions, with central deployments exceeding 40 GW. In the baseline case ($\kappa = 1.0$), a more balanced spatial distribution is achieved, as part of the capacity shifts toward low-carbon regions while maintaining overall economic efficiency. When $\kappa = 1.2$, the weight of carbon costs becomes dominant, resulting in a substantial reduction of capacity in carbon-intensive regions (approximately 28 GW) and a corresponding increase in allocations to low-carbon areas.

These results reveal that increasing the carbon cost factor induces a more spatially diversified deployment of data centers toward regions with lower carbon intensity, while a lower value favors concentration in cost-efficient yet carbon-intensive areas. This finding underscores the pivotal role of $\hbar$ in governing the trade-off between economic efficiency and environmental sustainability in large-scale data center planning.

\section{Conclusion and discussion}
This paper proposes a long-term joint planning framework for coordinating power systems and data centers from the perspective of system operators. The model integrates decisions on generation, storage, transmission, and data center siting and sizing, while accounting for both operational and embodied carbon emissions. The large-scale stochastic optimization is solved via an enhanced Benders decomposition algorithm and demonstrated on the PJM system, with all datasets publicly released on GitHub \cite{github-pjm-case}.

By implementing the model on the real-world PJM system, one of the world's largest and most critical electricity markets, this study moves beyond purely theoretical analysis and provides actionable insights for system operators and planners. The case study provides several key insights. First, the PJM system can accommodate up to 55 GW of peak data center demand, with DOM (Virginia) and ComEd (Illinois) identified as major hubs. Second, joint planning reduces total investment costs by 12.6\%, operational costs by 8.3\%, and carbon emissions by 5.6\% compared with independent planning. Third, incorporating embodied emissions promotes 25.5\% more renewable capacity and reduces operational emissions by 16.9\%, underscoring the importance of life-cycle carbon accounting.

Future research may extend this work in several directions. First, practical coordination mechanisms between system operators and data center investors should be developed, potentially through market-based incentives. Second, communication network modeling can be refined beyond fixed investment shares to capture realistic data flows. Third, it should be noted that the current model simplifies each PJM transmission zone as a single node, which does not capture intra-zonal congestion effects; future work could explore a more granular nodal representation to enhance the spatial accuracy of grid constraints. Fourth, the integration of advanced uncertainty modeling techniques, such as distinguishing between exogenous and endogenous sources of uncertainty \cite{ref-B} and incorporating option value theory \cite{ref-A}, could enhance the economic evaluation of flexible investments under uncertainty. Fifth, prediction modeling approaches from related domains, such as e-mobility grid integration \cite{ref-C}, could be adapted to improve demand forecasting for data center growth. Finally, incorporating discrete unit commitment into the planning framework would enhance operational realism, though more efficient formulations are needed to manage the associated computational challenges.

\appendix \section{Nomenclature}
\subsection{Abbreviations}
\begin{tabbing}
\hspace{\symbolwidth} \= \kill
\nomitem{CG}{Coal-fired generation unit}
\nomitem{DC}{Data center}
\nomitem{ES}{Energy storage}
\nomitem{NG}{Natural-gas generation unit}
\nomitem{PUE}{Power usage effectiveness}
\nomitem{PV}{Photovoltaic generation unit}
\nomitem{RES}{Renewable energy resource}
\nomitem{SOC}{State of charge}
\nomitem{SRV}{Servers in data center}
\nomitem{TC}{Data transmission facility}
\nomitem{TL}{Transmission lines}
\nomitem{WT}{Wind turbine unit}
\nomitem{Fix}{Fixed facilities of data center}
\end{tabbing}

\subsection{Variables}
\begin{tabbing}
\hspace{\symbolwidth} \= \kill
\nomitem{$C_{\mt{curt}}^{y}$}{Curtailment penalty for PV and WT of year $y$ (\$)}
\nomitem{$C_{\mt{DC}}^{y}$}{Investment cost of data centers of year $y$ (\$)}
\nomitem{$C_{\mt{ex}}^{y}$}{Energy exchange cost of the system of year $y$ (\$)}
\nomitem{$C_{\mt{gen}}^{y}$}{Total generation cost of all resources of year $y$ (\$)}
\nomitem{$C_{\mt{GR}}^{y}$}{Investment cost of generation resources of year $y$ (\$)}
\nomitem{$C_{\mt{om}}^{y}$}{Total maintenance cost of all resources of year $y$ (\$)}
\nomitem{$C_{\mt{TC}}^{y}$}{Investment for data transmission facilities of year $y$ (\$)}
\nomitem{$C_{\mt{TL}}^{y}$}{Investment cost of transmission lines of year $y$ (\$)}
\nomitem{$\mt{C}_{\mt{RT}}^{y}$}{Salvage value of retired equipments of year $y$ (\$)}
\nomitem{$E^{y,d,t,i}_{\mt{ES}}$}{State of charge of energy storage devices on node $i$ in year $y$, day $d$ and hour $t$ (MWh)}
\nomitem{$N_{k}^{y,i}$}{Number of installed units of resource $k$ at node $i$}
\nomitem{$P^{y,d,t,i}_{k}$}{Power output of resources at node $i$ in year $y$, day $d$ and hour $t$ (MW)}
\nomitem{$P^{y,d,t,i}_{\mt{BUY}}$}{Purchased energy of PJM from other ISOs at node $i$ in year $y$, day $d$ and hour $t$ (MW)}
\nomitem{$P^{y,d,t,i}_{\mt{DC}}$}{Power demand of data center at node $i$ in year $y$, day $d$ and hour $t$ (MW)}
\nomitem{$P^{y,d,t,i}_{\mt{k,cap}}$}{Available power output capability for resource of node $i$ in year $y$, day $d$ and hour $t$ (MW)}
\nomitem{$P^{y,d,t,i}_{\mt{SELL}}$}{Sold power of PJM to other ISOs from node $i$ in year $y$, day $d$, and hour $t$ (MW)}
\nomitem{$P^{y,d,t,l}_{\mt{TL}}$}{Power flow on transmission line $l$ in year $y$, day $d$ and hour $t$ (MW)}
\nomitem{$X^{y,d,t,i}_{\mt{SVR},m}$}{Number of active servers in a data center (rack)}
\nomitem{$Y_{\mt{DC}}^{y,d,t,i}$}{Computing workload served by data centers at node $i$ in year $y$, day $d$, hour $t$ (Requests)}
\nomitem{$Y_{\mt{TC}}^{y,d,t,ij}$}{Computing workload transferred from node $i$ to node $j$ in year $y$, day $d$, hour $t$ (Requests)}
\nomitem{$\bar{Y}_{ij}^{y}$}{Computing workload transfer capability from node $i$ to node $j$ in year $y$ (Requests)}
\nomitem{$\mathbf{x}$}{Abstract vector representing decision variables for the operational stage, including power outputs of all generation resources and data centers, as well as data workload decisions for servers}
\nomitem{$\mathbf{z}$}{Abstract vector representing decision variables for the planning stage, encompassing all installed capacity decisions, as well as newly added and retired capacity decisions}
\nomitem{$\mt{BN}_{\mt{SRV,m}}^{y,i}$}{Newly added number of $m$-type servers in year $y$ at node $i$ (rack)}
\nomitem{$\mt{CE}^{y}_{\mt{EMB}}$}{Total embodied carbon emissions of the system in year $y$ (short ton)}
\nomitem{$\mt{CE}^{y}_{\mt{OPR}}$}{Total operational carbon emissions of the system in year $y$ (short ton)}
\nomitem{$\mt{IC}_{k}^{y,i}$}{Installed capacity of system resources in year $y$ at node $i$ (MW)}
\nomitem{$\mt{IC}_{\mt{ES}}^{y,i}$}{Installed capacity of energy storage devices in year $y$ on node $i$ (MWh)}
\nomitem{$\mt{IN}_{\mt{DC}}^{y,i}$}{Installed number of all server types in the data center at node $i$ in year $y$ (rack)}
\nomitem{$\mt{IN}_{\mt{SRV,m}}^{y,i}$}{Installed number of $m$-type server in the data center at node $i$ in year $y$ (rack)}
\nomitem{$\mt{RN}_{\mt{SRV,m}}^{y,i}$}{Number of $m$-type servers retired in year $y$ at node $i$ (rack)}
\nomitem{$\mt{RV}$}{Total residual value of all resources at the end of the planning horizon (\$)}
\end{tabbing}

\subsection{Parameters}
\begin{tabbing}
    \hspace{\symbolwidth} \= \kill
    \nomitem{$\beta^{\min}_{k}$}{Capacity scaling factor representing the minimum output of resource $k$ (\%)}
    \nomitem{$\beta^{\max}_{k}$}{Capacity scaling factor representing the maximum output of resource $k$ (\%)}
    \nomitem{$\chi^{\mt{emb}}_{k}$}{Embedded carbon emission coefficient for resource $k$ (short ton/MW)}
    \nomitem{$\chi^{\mt{emb}}_{\mt{ES}}$}{Embodied carbon emission coefficient for energy storage devices (short ton/MWh)}
    \nomitem{$\chi^{\mt{emb}}_{\mt{Fix}}$}{Embodied carbon emission coefficient for fixed resources in data centers (short ton/rack)}
    \nomitem{$\chi^{\mt{emb}}_{\mt{SRV,m}}$}{Embodied carbon emission coefficient for $m$-type servers in data center (short ton/rack)}
    \nomitem{$\chi^{\mt{emb}}_{\mt{TC}}$}{Embodied carbon emission coefficient for data transmission facilities (short ton/km $\cdot$ Requests)}
    \nomitem{$\chi^{\mt{emb}}_{\mt{TL}}$}{Embodied carbon emission coefficient for transmission lines (short ton/MW $\cdot$ km)}
    \nomitem{$\chi^{\mt{gen}}_{k}$}{Generation carbon emission coefficient for resource $k$ (short ton/MWh)}
    \nomitem{$\delta^{\mt{data}}_{d,t}$}{Daily variation factor for computing workload demand in day $d$ and hour $t$ (\%)}
    \nomitem{$\delta^{\mt{load}}_{d,t}$}{Daily variation factor for electricity demand in day $d$ and hour $t$ (\%)}
    \nomitem{$\delta^{\mt{pv/wt}}_{d,t}$}{Daily variation factor for solar power and wind power outputs in day $d$ and hour $t$ (\%)}
    \nomitem{$\Delta t$}{Time interval (hour)}
    \nomitem{$\eta_{\mt{esc}}/\eta_{\mt{esd}}$}{Charging/Discharging efficiency for energy storage devices (\%)}
    \nomitem{$\gamma_{\mt{tech}}^{y}$}{Investment cost discount factor, representing the reduction in investment costs for GPU and other server equipments due to technological advancements (\%)}
    \nomitem{$\Gamma_{l,i}$}{Power transfer distribution factor (PTDF) from node $i$ to line $l$}
    
    \nomitem{$\hbar$}{Weighting factor that reflects the decision maker's preference toward carbon emission reduction (\$/short ton)}
    \nomitem{$\lambda^{\mt{data}}_{y}$}{Yearly growth factor for computing workload (\%)}
    \nomitem{$\lambda^{\mt{load}}_{y}$}{Yearly growth factor for electricity demand (\%)}
    \nomitem{$\lambda^{y}_{\mt{tech}}$}{Embodied carbon emission discount factor for GPU and other server equipments representing technological advancements (\%)}
    \nomitem{$\phi_{y,d,t,i}^{\mt{temp}}$}{Cooling efficiency scaling factor (\%)}
    \nomitem{$\psi_{\mt{SVR},m}$}{The equivalent processing capability of the $m$-type server (Requests/hour $\cdot$ rack)}
    \nomitem{$\sigma$}{Discount factor (\%)}
    \nomitem{$\varphi_{i}^{\mt{data}}$}{ Computing workload load share for node $i$ (\%)}
    \nomitem{$\varphi_{i}^{\mt{load}}$}{Electric load share for node $i$ (\%)}
    \nomitem{$\xi^{buy/sell}_{i,t}$}{Energy exchange price at node $i$ and hour $t$ (\$/MWh)}
    \nomitem{$\xi^{curt}_{k}$}{Curtailment cost coefficient for resource $k$ (\$/MWh)}
    \nomitem{$\xi_{k}^{gen}$}{Generation cost coefficient for resource $k$ (\$/MWh)}
    \nomitem{$\xi_{k}^{om}$}{Maintenance cost coefficient for resource $k$ (\$/MWh)}
    \nomitem{$\xi^{\mt{inv}}_{\mt{ES}}$}{Unit investment for energy storage units (\$/MWh)}
    \nomitem{$\xi^{\mt{inv}}_{\mt{Fix}}$}{Unit investment cost for fixed resources of a data center, e.g., land and material costs (\$/rack)}
    \nomitem{$\xi^{\mt{inv}}_{\mt{SRV,m}}$}{Unit investment cost for $m$-type server (\$/rack)}
    \nomitem{$\xi^{\mt{inv}}_{\mt{TC}}$}{Unit investment cost for data transmission facilities (\$/km $\cdot$ Requests)}
    \nomitem{$\xi^{\mt{inv}}_{\mt{TL}}$}{Unit investment for transmission lines (\$/km $\cdot$ MW)}
    \nomitem{$\xi^{\mt{inv}}_{k}$}{Unit investment cost for generation resource $k$ (\$/MW)}
    \nomitem{$\zeta_{s}$}{Stochastic scenario}
    \nomitem{$E^{\mt{unit}}_{ES}$}{Unit energy storage capacity (MWh)}
    \nomitem{$H_{l}$}{Length of transmission line $l$ (km)}
    \nomitem{$H_{ij}$}{Physical distance between node $i$ and $j$ (km)}
    \nomitem{$L_{k}$}{Lifetime of system resource $k$ (year)}
    \nomitem{$L_{\mt{SRV,m}}$}{Lifetime of $m$-type servers (year)}
    \nomitem{$R^{y}_{k}$}{Residual factor for resource $k$ installed in year $y$ (\%)}
    \nomitem{$R^{y}_{\mt{CG}}$}{Residual factor for coal-fired units retired in year $y$ (\%)}
    \nomitem{$R^{0}_{k}$}{Scrap value for resource $k$ at the end of its life (\%)}
    \nomitem{$R^{y}_{\mt{SRV,m}}$}{Residual value factor of $m$-type servers in year $y$ (\%)}
    \nomitem{$R^{0}_{\mt{SRV,m}}$}{Scrap value of $m$-type servers at the end of its life (\%)}
    \nomitem{$R^{\mt{up/down}}_{k}$}{Up/Down ramping rate for resources $k \in \{\mt{NG,CG}\}$ (\%)}
    \nomitem{$\bar{\mt{RT}}_{\mt{CG}}^{y}$}{Retired capacity limitation for CG units in year $y$ (MW)}
    \nomitem{$P^{\mt{unit}}_{k}$}{Unit generation capacity of resource $k$ (MW)}
    \nomitem{$P^{y,d,t,i}_{\mt{LOAD}}$}{Electric load of the system without data centers at node $i$ in year $y$, day $d$, hour $t$ (MW)}
    \nomitem{$P^{\max}_{\mt{EX}}$}{Maximum allowable power exchange (MW)}
    \nomitem{$P^{\mt{rate}}_{\mt{SVR,m}}$}{Rated power consumption of $m$-type server (MW/rack)}
    \nomitem{$P^{peak}_{\mt{LOAD}}$}{Peak power demand of the system (MW)}
    \nomitem{$\mt{PUE}_{m}$}{Power usage effectiveness for $m$-type server (\%)}
    \nomitem{$\bar{\mt{IC}}_{\mt{CG}}^{y}$}{Upper limit of capacity retirement of coal-fired generation units of year $y$ (MW)}
    \nomitem{$\bar{\mt{IC}}_{k}^{y,i}$}{Maximum installed capacity limitation of resource $k$ in year $y$ at node $i$ (MW)}
    \nomitem{$\underline{\mt{IC}}_{k}^{y,i}$}{Minimum installed capacity requirement of resource $k$ in year $y$ at node $i$ (MW)}
    \nomitem{$\bar{\mt{IC}}_{\mt{ES}}^{y,i}$}{Maximum installed capacity limitation of energy storage devices in year $y$ at node $i$ (MWh)}
    \nomitem{$\underline{\mt{IC}}_{\mt{ES}}^{y,i}$}{Minimum installed capacity requirement of energy storage devices in year $y$ on node $i$ (MWh)}
    \nomitem{$Y_{\mt{LOAD}}^{y,d,t,i}$}{Computing workload demand on node $i$ in year $y$, day $d$, hour $t$ (Requests)}
    \nomitem{$Y^{peak}_{\mt{LOAD}}$}{Peak computing workload of the system (Requests)}
    \nomitem{$Y_{ij}^{0}$}{Existing data transmission resources from node $i$ to node $j$ (Requests)}
    \nomitem{$Z_{ij}$}{Binary parameter indicating whether data transmission facilities should be built between node $i$ and $j$}
\end{tabbing}

\bibliographystyle{elsarticle-num}
\bibliography{cas-refs}

\end{document}